\documentclass{jkas}
\usepackage{longtable}
\usepackage{threeparttable}
\usepackage{siunitx}
\def\year{2023} 
\def\volume{56} 
\def\issue{1} 
\def\beginpage{1} 
\def\received{---} 
\def\accepted{---} 
\def\published{---} 
\date{Received \received; Accepted \accepted; Published \published}

\newcommand\ion[2]{{#1}\,{\sc #2}} 

\title{%
Constructing a Hydrogen Line Library for Ly$\alpha$ Emitters at Low Redshifts ($z\lesssim0.4$): Estimating Dust Extinction and Assessing Paschen line Detectability with SPHEREx
}

\author[1]{Junho Song}{0009-0006-0197-890X}
\author[2]{Hyunmi Song}{0000-0002-4362-4070}
\author[3]{Hyunjin Shim}{0000-0002-4179-2628}

\affil[1]{Department of Earth Environmental and Space Convergence Sciences, Chungnam National University, 99 Daehak-ro, Yuseong-gu, Daejeon 34134, Republic of Korea}
\affil[2]{Department of Astronomy and Space Science, Chungnam National University, 99 Daehak-ro, Yuseong-gu, Daejeon 34134, Republic of Korea}
\affil[3]{Department of Earth Science Education, Kyungpook National University, 80 Daehak-ro, Buk-gu, Daegu 41566, Republic of Korea}

\def\corrauthor{%
Hyunmi Song, Hyunjin Shim \email{hmsong@cnu.ac.kr, hjshim@knu.ac.kr}
}

\def\runningauthor{%
Song et al.
}

\def\runningtitle{%
Constructing a Hydrogen Line Library for ${\rm Ly}\alpha$ Emitters at Low Redshifts
}

\def\keywords{%
surveys --- infrared: galaxies --- ISM: dust, extinction --- galaxies: star formation
}

\def\abstracttext{%
Hydrogen recombination lines, one of the strongest emission lines from star-forming galaxies, are used to probe the early Universe as indicators of star formation rates and ionizing photon production rates. 
Ratios between different recombination lines provide clues to estimate dust attenuation, in addition to the \ion{H}{II} region physical diagnostics. 
To prepare for future near-infrared spectral surveys and optical narrow-band imaging surveys aiming for hydrogen lines in different redshifts, we construct hydrogen recombination line libraries by compiling data from published literature. 
Our compilation includes 260 galaxies mostly at $z\lesssim0.4$, of which at least one hydrogen recombination line is observed. 
The specific hydrogen emission lines under investigation encompass three Balmer lines (${\rm H}\alpha$, ${\rm H}\beta$ and ${\rm H}\gamma$) and ${\rm Ly}\alpha$ line. 
Of the compiled galaxies, we estimated dust extinction for 85 galaxies that have detections of at least two Balmer lines, by assuming the Case B recombination and Calzetti extinction curve, based on which ${\rm Ly}\alpha$ escape fraction is also inferred.
For each of the 14 galaxies with detections of all three Balmer lines, we optimized the extinction curve so that it yields consistent $E(B-V)$ values across the three possible Balmer line ratios, showing the variety in the slope of the dust attenuation curve.
We further evaluated the feasibility of studying hydrogen emission line-selected galaxies with future spectral surveys such as SPHEREx by modeling the Paschen lines of our galaxy sample.
}

\begin{document}

\jkashead 


\section{Introduction}\label{sec:1intro}

Hydrogen recombination lines are one of the most well-studied star formation rate (SFR) indicators \citep[e.g.][]{Kennicutt1998, KennicuttEvans2012}.
While other continuum-based SFR indicators, such as the ultraviolet (UV) continuum from massive young stars and re-processed infrared (IR) emission by dust, probe longer time scales of star formation (e.g., hundreds of Myrs for B-type stars in the case of UV, and a prolonged Gyr scale including A-type stars in the case of IR),
the line emission from the ionized gas around massive stars is tracer of shorter time scale of star formation ($\sim10$\,Myrs) as the primary contributors to this line emission are stars with masses of 30-40 M$_\odot$ \citep{KennicuttEvans2012}.
The fact that different SFR indicators probe different time scales of star formation leads us to use the ratio between the continuum and line emissions as an indicator of star formation history, i.e.,
measurement of `burstiness' of star formation \citep[e.g.][]{Guo2016, Byun2021}.

Since line emission is a probe of star formation, observational strategies are developed to use the existence of strong emission lines to sample vigorous star-forming galaxies at different redshifts, using blind spectroscopic surveys (frequently using slitless prism/grism) or narrow-band imaging surveys where strong line emission elevates the magnitudes measured in the specific filter \citep{Ouchi2008, Deharveng2008Galex, Cowie2010, Ouchi2018, HayashiSobral2013, Cochrane2017, Sun2018, Cheng2020, Khostovan2024, Cooper2023, HerenzEdmund2019, ODIN}.
Galaxies selected this way are generally termed `line emitters' and are widely used to investigate galaxy evolution through cosmic history, by allowing the estimation of the cosmic SFR density as a function of redshift.
Especially ${\rm Ly}\alpha$ emitters (LAEs), selected using the rest-frame UV resonance line of hydrogen, may probe the highest redshift Universe and are considered to have played a significant role in producing the ionizing photon budget that reionized the Universe. 

However, rest-frame UV and optical wavelengths are susceptible to dust attenuation; therefore, the SFR estimated from indicators in these wavelengths needs to be corrected for dust attenuation.
In the case of the UV continuum, methods have been proposed to use the UV slope $\beta$ \citep{Meurer1999, Calzetti2000, Takeuhi2012, Casey2014, Hamed2023} to correct for dust attenuation. 
This process is complicated as it heavily depends on the use of different attenuation laws \citep[e.g.][]{Salimnaraynan2020}.
Other strategies, such as combining UV and IR SFR to represent the total value of SFR \citep[e.g.][]{Wuyts2011, Whitaker2014, Brown2017}, have been suggested, yet they also suffer from uncertainties in the modeling the IR emission spectral energy distribution (SED).
On the other hand, the line ratios between different hydrogen recombination lines have been suggested as a useful indicator of dust attenuation, as they are insensitive to the metallicity, gas temperature, and density of the ionized gas \citep{OsterbrockFerland2006}.
The deviation of the measured Balmer decrement, i.e., ${\rm H}\alpha$/${\rm H}\beta$ line ratio, from the theoretical values assuming the Case B recombination (e.g., 2.86 at $T=10,000\,{\rm K}$ and $n_e=100\,{\rm cm^{-3}}$), is easily converted to gas attenuation (with the given attenuation law) and stellar attenuation $A_V$ is widely used \citep{CalzettiKinney1994, PettiniShapley2001, OsterbrockFerland2006}.

Nevertheless, recent near-infrared observations suggest that such an analysis (using the Balmer decrement to estimate dust attenuation) can still be inaccurate under certain conditions, such as when \ion{H}{II} regions are optically thick.
Studies using MIR hydrogen lines have suggested that the dust attenuation measured by MIR-to-optical hydrogen line ratios (such as ${\rm Pa}\beta/{\rm H}\alpha$) can differ from that measured by the Balmer decrement \citep[e.g.][]{Cleri2022, Payne2018, Yano2021, Reddy2023, Pastrav2023}.
Because Hydrogen recombination lines in the MIR--such as the Paschen and Brackett series (1--5\,$\mu {\rm m}$)--are less affected by dust, they offer more reliable, attenuation-independent estimates of SFRs.
They can also provide insights into the true dust content, even in extreme environments like heavily obscured ultra-luminous infrared galaxies \citep[ULIRGs;][]{Imanishi2010, KennicuttEvans2012, Yano2021} and dust-obscured quasars \citep{Kim2023quasar}. 
Furthermore, using multiple hydrogen line ratios enables more accurate constraints on the shape of the dust attenuation curve, rather than relying on a fixed or assumed curve \citep{Prescott2022, JiYan2023, LinYan2024}.
Therefore, the Paschen and Brackett series hold strong potential for improving SFR and dust attenuation estimates \citep{CleriTrump2022, Reddy2023}.
The Spectro-Photometer for the History of the Universe, Epoch of Reionization, and ices Explorer \citep[SPHEREx;][]{SPHEREX2020}, the ongoing spectrophotometric all-sky survey in the 1--5\,$\mu \rm{m}$ range, will enable a blind search for Paschen/Brackett line-selected sources in the nearby and low-redshift Universe, complementing existing optical line surveys. 
In the case of LAEs with UV observations, it would be possible to combine multi-wavelength data from UV to IR to better understand their physical conditions, which help constrain the ionizing photon escape fraction.

In this study, we aim to assess the feasibility of using multiple Hydrogen recombination lines--both those already observed and those expected to be observed--to constrain dust attenuation, such as the color excess $E(B-V)$ and the slope of the attenuation curve. 
As a first step, we compiled a sample of known line emitters from the literature, focusing on LAEs at relatively low redshifts ($z\lesssim0.4$). 
For these galaxies, we estimated their dust properties using available Balmer lines measured in previous studies. 
Hydrogen recombination lines in the near-infrared (NIR), which would be obtained by SPHEREx, are expected to be more useful in studying dust properties than Balmer lines, since dust attenuation is weaker at longer wavelengths.
To explore this potential, we modeled the expected Paschen line fluxes using available information and assessed their detectability with SPHEREx.

This paper is organized as follows. Our data compilation is described in Section~\ref{sec:2data}.
Dust attenuation, ${\rm Ly}\alpha$ escape fraction, and the slope of the attenuation curve for our sample galaxies are estimated from the compiled Balmer line measurements, as described in Section~\ref{sec:3}.
For the construction of a more complete hydrogen line library, the detectability of additional hydrogen lines with SPHEREx is examined in Section~\ref{sec:4}, and a summary is provided in Section~\ref{sec:5summary}.
Throughout this paper, we use Vega magnitudes and assume a flat $\Lambda$CDM cosmology with $H_0=67.4\,{\rm km\,s^{-1}\,Mpc^{-1}}$ and $\Omega_m=0.315$ \citep{PlankCollaboration2020}.

\section{Data}\label{sec:2data}

\begin{figure*}
    \centering
    \includegraphics[width=\linewidth]{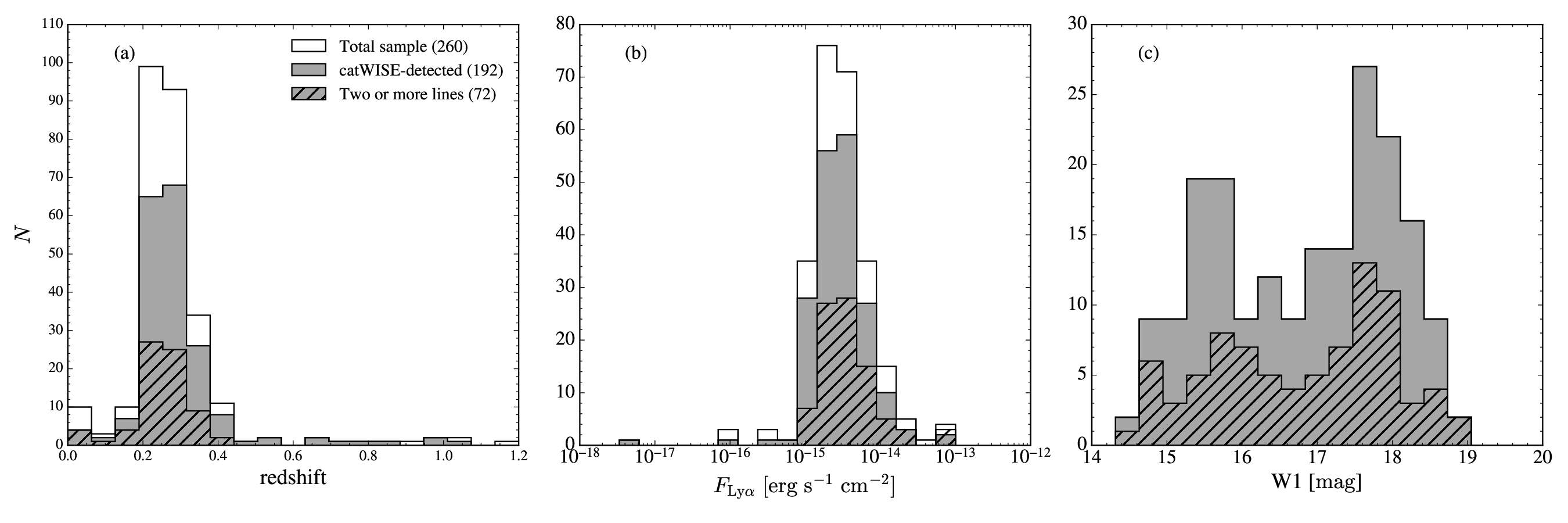}
    \caption{Distributions of redshift, ${\rm Ly}\alpha$ flux ($F_{\rm Ly\alpha}$), and catWISE W1 magnitude for our sample galaxies. The empty histogram represents 260 galaxies with spectroscopic redshifts; the filled histogram is a subset of these, showing 192 galaxies that are cross-matched with the catWISE source catalog; and the hatched histogram represents the 72  catWISE-detected galaxies out of the 85 that have at least two Balmer emission line measurements.
}
    \label{fig:1redshift}
\end{figure*}

We constructed a literature-based sample of low-redshift LAEs ($z\lesssim0.4$) from six previous studies: \citet[O14]{Ostlin2014}, \citet[S09]{Scarlata2009}, \citet[C11]{Cowie2011}, \citet[F11]{Finkelstein2011}, \citet[A14]{Atek2014}, and \citet[W17]{Wold2017}.
Most of these studies are ultimately rooted in the GALEX low-redshift LAE samples \citep[][]{Deharveng2008Galex, Cowie2010, Wold2017}, although one subsample (LARS; O14) was selected separately.
In this context, ``GALEX-selected'' refers specifically to galaxies identified through GALEX UV slitless spectroscopy. 
This emphasis on GALEX-based samples reflects the fact that GALEX UV slitless spectroscopy has provided the primary wide-area datasets in which ${\rm Ly}\alpha$ emission is directly identified at low redshift. 
As a result, it serves as a natural foundation for assembling a low-$z$ LAE compilation.

Although most of the sample shares a common origin in GALEX spectroscopic LAE catalogs, the merged dataset is not statistically homogeneous. 
However, this heterogeneity does not compromise the main objective of this study. 
Our primary goal is not to derive unbiased population statistics or to characterize the intrinsic demographics of low-$z$ LAEs. Rather, we aim to assemble a hydrogen-line library from the literature and to extend it by assessing the observability of additional hydrogen emission lines, particularly the Paschen series in the context of SPHEREx.
For this feasibility-oriented purpose, the key requirement is the availability of published hydrogen-line measurements and continuum constraints over the relevant wavelength range, rather than a globally uniform parent-sample selection.

Nevertheless, it is important to note that this inhomogeneity arises from two main factors.
First, the parent GALEX LAE catalogs themselves differ in their extraction strategies and AGN filtering, especially between the earlier continuum-preselected samples and the later continuum-blind sample. 
\citet{Deharveng2008Galex} and \citet{Cowie2010} relied on continuum-based spectral extraction and UV spectroscopic AGN rejection, whereas \citet{Wold2017} performed a continuum-blind emission-line search using reconstructed GALEX data cubes together with multi-wavelength AGN identification. As a result, the underlying LAE populations sampled by these catalogs are not identical, with the latter being more sensitive to high-EW systems with faint UV continua.

Second, the individual follow-up studies applied additional criteria tailored to their own observational goals, such as restricting the redshift range to ensure coverage of specific rest-frame optical lines, selecting particular survey fields, imposing emission-line EW thresholds, or relying on the availability of ancillary spectroscopy. 
14 LAEs at $z<0.18$ from O14 were selected through GALEX-SDSS cross-matching with $\mathrm{EW}(\mathrm{H}\alpha)>100\,\unit{\angstrom}$.
29 LAEs at $z\sim0.28$ from S09 are GALEX LAEs restricted to sources in several Palomar-accessible fields at redshifts where ${\rm H}\alpha$ and ${\rm H}\beta$ fall within the optical spectral range.
96 LAEs at $z=0.195$--$0.44$ (plus 8 at $z=0.65$--$1.25$) from C11 are GALEX LAEs with $\mathrm{EW}(\mathrm{Ly}\alpha)\gtrsim20\,\unit{\angstrom}$.
12 LAEs at $z\sim0.32$ from F11 are GALEX LAEs selected for spectroscopic follow-up in the EGS field.
24 LAEs at $z\sim0.39$ from A14 are GALEX LAEs in the CDFS and ELAIS-S1 fields with available optical spectroscopy. 
171 LAEs at $z\sim0.29$ from W17 are faint continuum GALEX LAEs confirmed from optical spectroscopic follow-up. ${\rm H}\alpha$ and ${\rm H}\beta$ lines are identified in the optical spectra, but the line fluxes are not published and thus only ${\rm Ly}\alpha$ fluxes are available for our compilation.
 
By cross-matching across the six LAEs samples, we compiled a total of 289 LAEs. 
After excluding sources without any hydrogen line flux information provided, 260 LAEs with ${\rm Ly}\alpha$ line flux detections remain in our sample. 
Among them, 14 have all three Balmer lines--${\rm H}\alpha$, ${\rm H}\beta$, and ${\rm H}\gamma$--available, and an additional 71 have both ${\rm H}\alpha$ and ${\rm H}\beta$ lines.
12 LAEs have ${\rm H}\alpha$ as the only available Balmer line (in addition to ${\rm Ly}\alpha$), and the remaining 163 have detections of the ${\rm Ly}\alpha$ line only.

We also cross-matched our compiled sample of 260 LAEs with the catWISE and 2MASS source catalogs, resulting in 192 matches with catWISE/W1, among which three LAEs were also matched with 2MASS.
For these 192 LAEs, the W1 flux is used as a continuum level in our SED modeling to assess the detectability of the Paschen lines (Section~\ref{sec:41spherex}). 
The distributions of redshifts, ${\rm Ly}\alpha$ fluxes, and W1 magnitudes are shown in Figure~\ref{fig:1redshift}.

\section{Dust Attenuation Properties from Balmer Lines}\label{sec:3}
\subsection{Dust attenuation and Ly$\alpha$ escape fraction}\label{sec:31dust}
We estimated the dust attenuation of LAEs with at least two detected Balmer lines. 
The deviation of the observed flux ratio between two hydrogen lines ($({\rm H}_i/{\rm H}_j)_{\rm obs}$\footnote{If no subscript is indicated, it means an observed flux ratio hereafter.}) from its intrinsic value ($({\rm H}_i/{\rm H}_j)_{\rm int}$), referred to as the Balmer decrement, can be translated into dust attenuation.
Using the relation $A(\lambda)=\kappa(\lambda)E(B-V)$, where $A(\lambda)$ is the dust attenuation, $\kappa(\lambda)$ is a given attenuation curve, and $E(B-V)$ is the color excess ($A(B)-A(V)$), we can then derive
\begin{equation}
E(B-V) = 
\frac{-2.5}{\kappa ({\rm H}_j)-\kappa({\rm H}_i)} \times 
\log_{10}\left[\frac{({\rm H}_i/{\rm H}_j)_{\rm int}}{({\rm H}_i/{\rm H}_j)_{\rm obs}}\right].
\label{equ:1ebv}
\end{equation}
The intrinsic ratios are determined under the Case B recombination condition, depending on the gas temperature and electron density. 
For typical interstellar medium conditions ($T=10^4\,{\rm K}$ and $n_e=10^2\,{\rm cm^{-3}}$), 
the intrinsic ratios are $({\rm H}\alpha/{\rm H}\beta)_{\rm int}=2.86$, 
$({\rm H}\gamma/{\rm H}\beta)_{\rm int}=0.469$, and
$({\rm H}\alpha/{\rm H}\gamma)_{\rm int}=6.09$. 
Since our targets are star-forming galaxies, we adopt the \citet{Calzetti2000} attenuation curve, which is widely used for star-forming galaxies:
\begin{equation}
\kappa(\lambda)=
\begin{aligned}
&\begin{cases}
2.658(-1.857+1.040\chi)+ R_V, \\ 
\qquad (0.63\,\mu\mathrm{m} \le \lambda \le 2.20\,\mu\mathrm{m}) \\[4pt]
2.659(-2.156+1.509\chi-0.198\chi^2+0.011\chi^3)+R_V, \\
\qquad (0.12\,\mu\mathrm{m} \le \lambda \le 0.63\,\mu\mathrm{m})
\end{cases}
\end{aligned}
\label{equ:2kappa}
\end{equation}
where $\chi=1/\lambda$ is the wave number and $R_V$ is the ratio of the total to selective attenuation defined as $A(V)/E(B-V)$, which is 4.05 for the Calzetti attenuation curve.

\begin{figure}
    \centering
    \includegraphics[width=\linewidth]{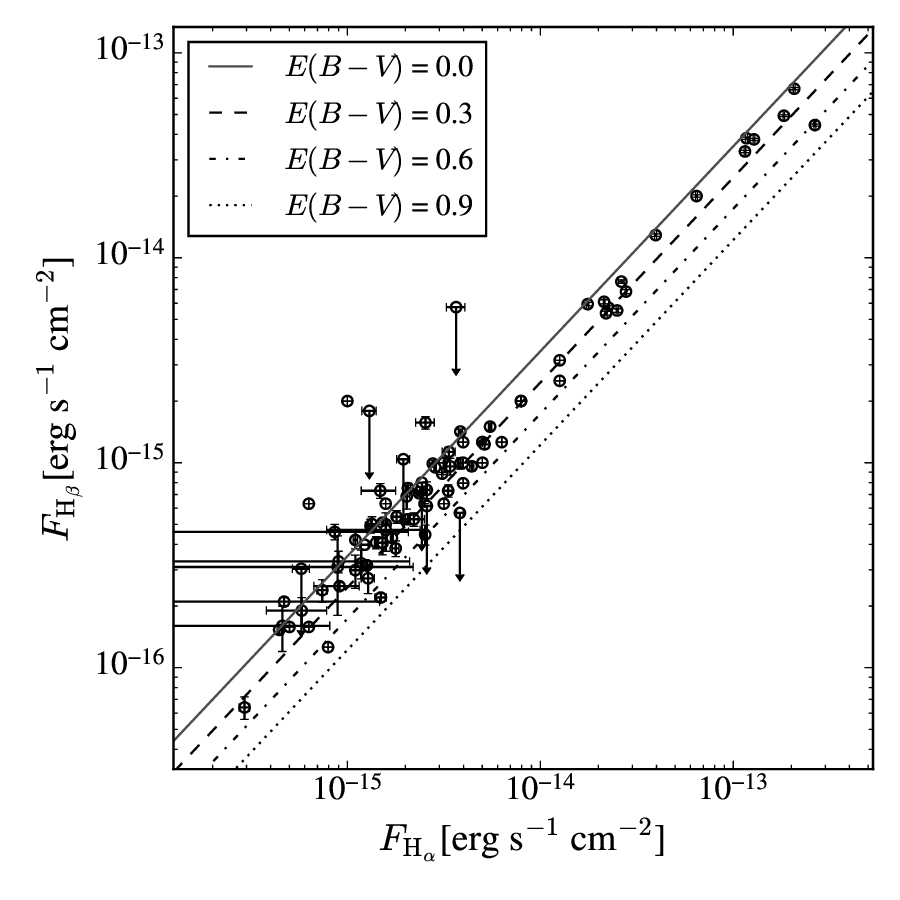}
    \caption{${\rm H}\beta$ line flux versus ${\rm H}\alpha$ line flux for 92 galaxies with ${\rm H}\alpha$ and ${\rm H}\beta$ flux measurements with their uncertainties (85 with direct detections and 7 with ${\rm H}\beta$ upper limits). Data points with downward arrows denote the 7 galaxies with upper limit for the ${\rm H}\beta$ fluxes.
    $E(B-V)$ values can be derived from the observed ${\rm H}\alpha/{\rm H}\beta$ flux ratio relative to its intrinsic ratio (the Balmer decrement; see the text and Equation (\ref{equ:1ebv})). The expected relations for four different $E(B-V)$ values are shown (0.0, 0.3, 0.6, and 0.9 with solid, dashed, dot-dashed, and dotted lines, respectively). Most galaxies have $E(B-V)<0.3$.}
    \label{fig:2hahb}
\end{figure}

Figure \ref{fig:2hahb} shows the observed fluxes of ${\rm H}\alpha$ and ${\rm H}\beta$ for 92 galaxies with the expected relations for varying $E(B-V)$ (denoted by dotted lines; from top to bottom corresponding to increasing $E(B-V)$). 
Most galaxies are distributed between $E(B-V)=0$ and 0.3.
A subset of galaxies has formally negative $E(B-V)$ values, i.e., they lie above the $E(B-V)=0$ line (${\rm H}\beta > {\rm H}\alpha/2.86$).
Recomputing $E(B-V)$ with alternative attenuation curves does not remove this negative tail, indicating that it is not primarily driven by the assumed attenuation law.
Provided that the flux measurements are reliable, these cases may suggest that the adopted Case B recombination assumption with $T=10^4\,{\rm K}$ and $n_e=10^2\,{\rm cm^{-3}}$ may not be appropriate \citep[e.g.,][]{Scarlate2024}. 
However, it more likely reflects measurement uncertainties such as uncertainties in the underlying stellar Balmer absorption corrections, which can easily scatter the observed line ratios of these galaxies below 2.86. 
Additional contributions may arise from flux-calibration uncertainties and low signal-to-noise measurements, particularly for the weaker H$\beta$ line.
The flux uncertainties adopted from the literature are unlikely to fully capture all systematic effects. Consequently, the scatter in the observed Balmer decrements is likely underestimated when considering only the published flux errors.
Nevertheless, in the absence of the original spectra and a homogeneous way to estimate these additional uncertainties, we retain the published flux errors in the subsequent analysis.

\begin{figure}
    \includegraphics[width=\linewidth]{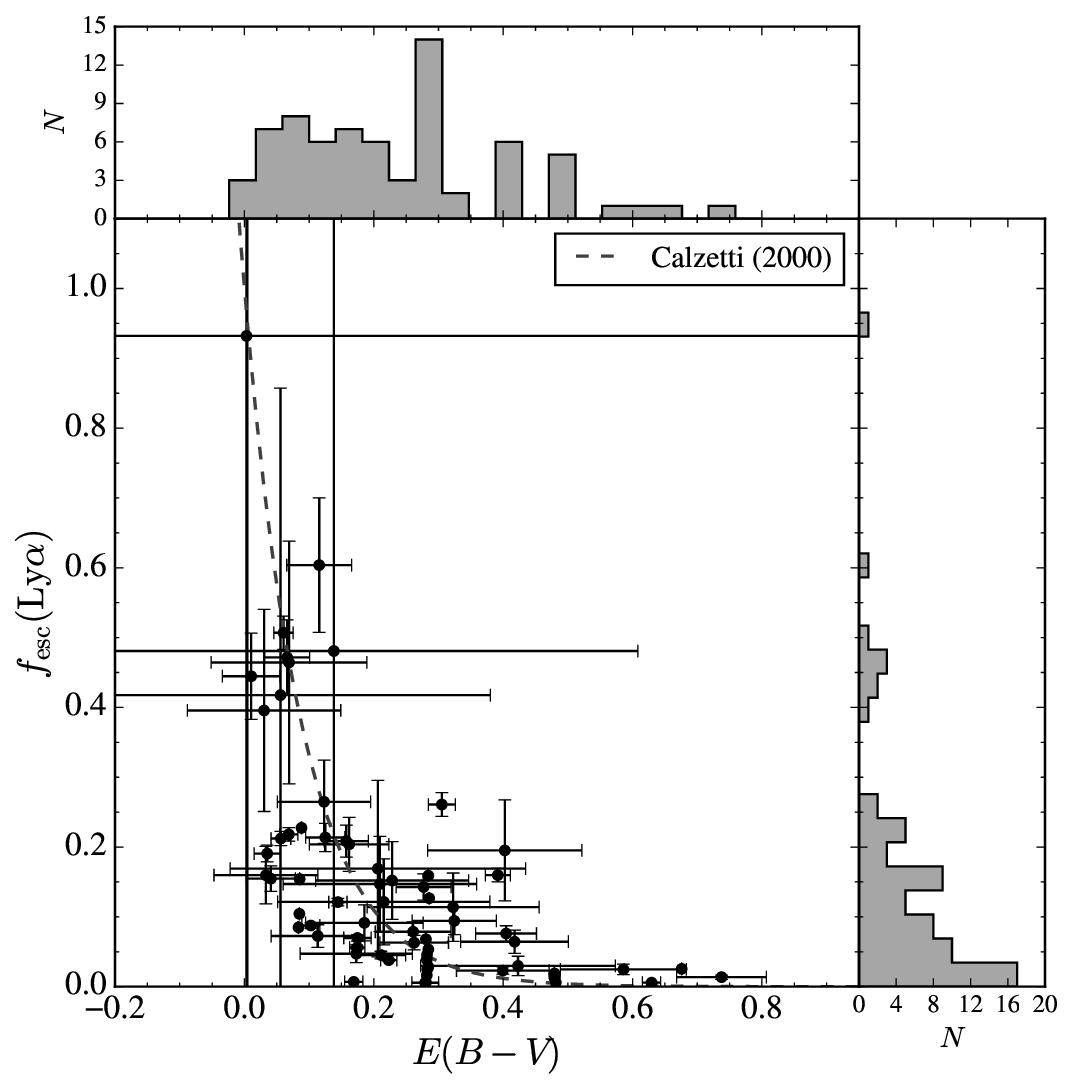}
    \caption{${\rm Ly}\alpha$ escape fraction ($f_{\rm esc}({\rm Ly}\alpha)$) as a function of $E(B-V)$. Here, $E(B-V)$ is derived from the Balmer decrement, and $f_{\rm esc}({\rm Ly}\alpha)$ is calculated as the ratio of the observed ${\rm Ly}\alpha$ flux to the intrinsic ${\rm Ly}\alpha$ flux. The intrinsic flux is inferred from the dust-corrected ${\rm H}\alpha$ flux assuming Case B recombination ($T=10^4\,{\rm K}$, $n_e=10^2\,{\rm cm^{-3}}$). Since individual errors for the observed ${\rm Ly}\alpha$ fluxes were not provided, we adopted a uniform uncertainty of $4\times10^{-16}\,\mathrm{erg\,s^{-1}\,cm^{-2}}$ for all galaxies, based on the typical measurement error of the GALEX spectroscopic survey \citep{Deharveng2008Galex}. This uncertainty was then propagated to determine the final error on $f_{\rm esc}({\rm Ly}\alpha)$. The escape fraction decreases with increasing $E(B-V)$, as expected. For comparison, the dotted line shows the prediction from \citet{Calzetti2000} assuming a simple dust screen and no ${\rm Ly}\alpha$ resonance scattering.}
    \label{fig:3escape}
\end{figure}

For the remaining 71 galaxies with $E(B-V) > 0$, we estimated $E(B-V)$ values from Equation (\ref{equ:1ebv}). The observed ${\rm H}\alpha$ fluxes were then corrected for dust attenuation to obtain their intrinsic values. Finally, we inferred the intrinsic ${\rm Ly}\alpha$ fluxes under the Case B recombination assumption (i.e., $({\rm Ly}\alpha/{\rm H}\alpha)_{{\rm int}} = 8.7$).
The ${\rm Ly}\alpha$ escape fraction, $f_{\rm esc}(\rm{Ly}\alpha)$, was then calculated as the ratio of the observed to intrinsic ${\rm Ly}\alpha$ flux.
Since individual uncertainties for the observed ${\rm Ly}\alpha$ fluxes were not provided, we adopted the uniform 1$\sigma$ line-flux precision of the GALEX spectroscopic survey \citep[$4\times10^{-16}\,\mathrm{erg\,s^{-1}\,cm^{-2}}$; ][]{Deharveng2008Galex} for all galaxies.
This uncertainty is then propagated into the uncertainty of $f_{\rm esc}(\rm{Ly}\alpha)$, along with that of $E(B-V)$.
Figure \ref{fig:3escape} shows the $E(B-V)$ and $f_{\rm esc}({\rm Ly}\alpha)$ distribution of 71 galaxies, which exhibits the expected anti-correlation; higher dust content (larger $E(B-V)$) leads to a lower probability of ${\rm Ly}\alpha$ photon escape.
The dotted curve is the expected relation when the resonance scattering of ${\rm Ly}\alpha$ is ignored and a simple dust screen is assumed \citep[see][]{Calzetti2000}.
As resonance scattering prolongs the path of ${\rm Ly}\alpha$ photons and increases their exposure to dust, causing stronger attenuation, the dotted curve can be regarded as an upper limit on $f_{\rm esc}({\rm Ly}\alpha)$ in a simple dust screen geometry.
Galaxies below the curve are likely affected by resonance scattering, whereas those above it are likely explained by a clumpy geometry in which $\rm{Ly}\alpha$ photons can escape with lower exposure to dust.
Most galaxies exhibit low escape fractions ($f_{\rm esc}({\rm Ly}\alpha) \lesssim 0.3$), and only a few reach $f_{\rm esc}({\rm Ly}\alpha) \gtrsim 0.4$. 
Although one galaxy shows $f_{\rm esc}({\rm Ly}\alpha) >0.9$, this estimation is not particularly significant due to the large uncertainties in $E(B-V)$; a few other galaxies also have large uncertainties, primarily originating from substantial ${\rm H}\alpha$ flux measurement errors.

In the analysis above, we adopted the Calzetti attenuation law as a fiducial prescription to derive $E(B-V)$ and the intrinsic ${\rm H}\alpha$ flux. 
This choice is practical because it is widely adopted for star-forming galaxies \citep[e.g.,][]{Calzetti2000,Salimnaraynan2020} and facilitates comparison with previous studies.
However, it is not necessarily universally applicable to LAEs, whose attenuation curves may show significant object-to-object variation. 
To assess the robustness of our results against this assumption, we also recomputed $E(B-V)$ using alternative attenuation prescriptions, including an SMC-like curve and other attenuation curves explored in Section~\ref{sec:32constraints}. 
We refer the reader to Section~\ref{sec:32constraints} for a discussion of the attenuation-curve diversity and for a quantitative comparison.

\subsection{Constraints on the shape of the dust attenuation curve}\label{sec:32constraints}
Multiple line combinations can be used to constrain not only $E(B-V)$ but also the slope of the attenuation curves. 
In particular, for the 14 galaxies in our sample with all three Balmer lines (${\rm H}\alpha$, ${\rm H}\beta$, and ${\rm H}\gamma$) detected, we are able to constrain the slope parameter $\delta$ of the attenuation law
\begin{equation}
A(\lambda) = A_V \left(\frac{\lambda}{\lambda_V}\right)^\delta
\label{equ:3alambda}
\end{equation}
by requiring that a single $E(B-V)$ value simultaneously reproduces the three line fluxes. 
By applying Equation (\ref{equ:3alambda}) to two independent line flux ratios and combining the resulting expressions, we obtain the following relation:
\begin{equation}
\begin{aligned}
    & \log\left(\frac{F_{\rm H\beta}}{F_{\rm H\gamma}}\right)-\log\left(\frac{F_{\rm H\beta}}{F_{\rm H\gamma}}\right)_{\rm int} \\
    & = \left[\log\left(\frac{F_{{\rm H}\alpha}}{F_{{\rm H}\beta}}\right)-\log\left(\frac{F_{{\rm H}\alpha}}{F_{{\rm H}\beta}}\right)_{\rm int}\right] \times
    \frac{\left(\lambda_{{\rm H}\beta} \right)^\delta - \left(\lambda_{{\rm H}\gamma} \right)^\delta}
    {\left(\lambda_{{\rm H}\alpha} \right)^\delta - \left(\lambda_{{\rm H}\beta} \right)^\delta}.
\end{aligned}
\label{equ:4fhbfha}
\end{equation}

\begin{figure}
    \centering
    \includegraphics[width=\linewidth]{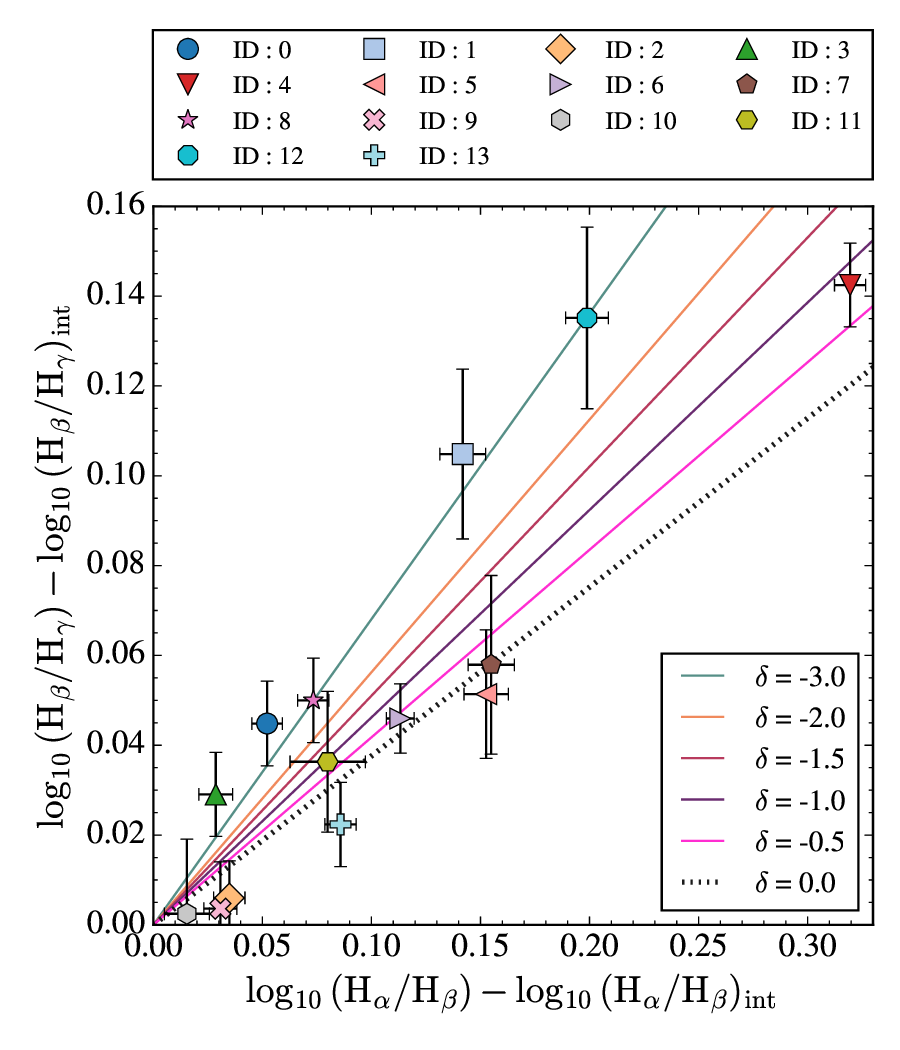}
    \caption{${\rm H}\alpha/{\rm H}\beta$ versus ${\rm H}\beta/{\rm H}\gamma$ flux ratios for the 14 galaxies with three Balmer line detections. Each ratio is normalized by its assumed intrinsic value (i.e., 2.86 for ${\rm H}\alpha/{\rm H}\beta$ and 2.13 for ${\rm H}\beta/{\rm H}\gamma$). The expected relation between the two ratios is shown for varying $\delta$ values, the slope of the attenuation curve (see Equation~(\ref{equ:4fhbfha})). A Calzetti-like attenuation curve slope is $\delta =-1.0$. Different $\delta$ values are required for individual galaxies, highlighting the advantage of utilizing more than two hydrogen lines. The dotted line denotes the boundary of the physically prohibited region, as it would require $\delta>0$ (i.e., larger attenuation at longer wavelengths). Six galaxies are located in this prohibited region: three near the origin likely have little dust content, while the adopted intrinsic line ratios may not be applicable to the other three.}
    \label{fig:4fhafhb}
\end{figure}

Figure \ref{fig:4fhafhb} shows the flux ratios ${\rm H}\beta/{\rm H}\gamma$ and ${\rm H}\alpha/{\rm H}\beta$, each normalized by their respective intrinsic values, together with the expected relations for varying $\delta$ values (colored solid lines) given by Equation (\ref{equ:4fhbfha}). 
The black dotted line indicates a flat (wavelength-independent) attenuation curve, which defines the boundary of a physically prohibited region. 
These results demonstrate that different $\delta$ values are required for individual galaxies to consistently explain the two observed flux ratios.
Six galaxies, including three near the origin with very little dust attenuation, lie in the prohibited region, suggesting that their intrinsic flux ratios are different from those adopted here (i.e., 2.86 and 2.13 for $({\rm H}\alpha/{\rm H}\beta)_{\rm int}$ and $({\rm H}\beta/{\rm H}\gamma)_{\rm int}$, respectively).
We note that intrinsic line ratios can vary depending on the physical conditions of a galaxy. Observational studies of AGNs have confirmed that such variations do occur \citep[e.g.,][]{Kim2010,KimIm2018}. While these variations could in principle affect the measured $E(B-V)$ values, their impact might be small compared to other sources of $E(B-V)$ measurement uncertainty--such as those arising from the choice of extinction curve or other observational factors--which are on the order of $\sim0.2$ \citep{Kim2018ebv}.

\begin{figure}
    \centering
    \includegraphics[width=\linewidth]{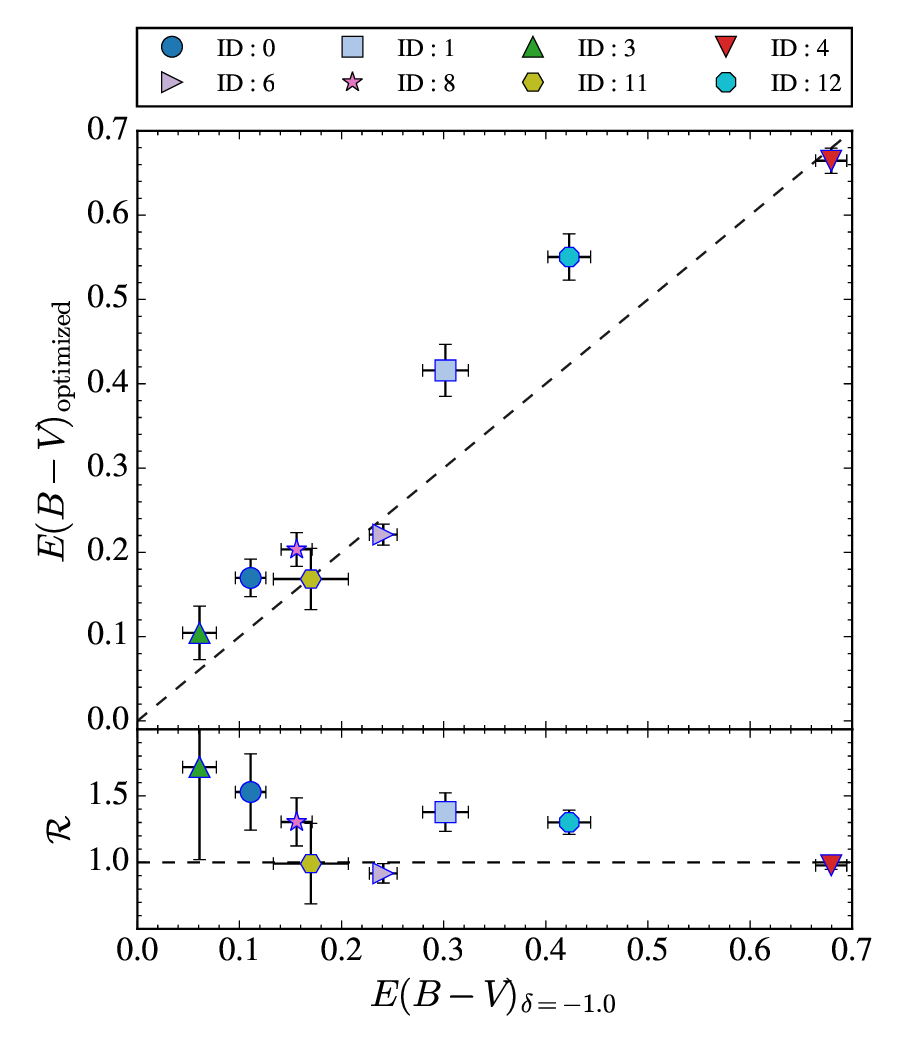}
    \caption{$E(B-V)$ values assuming the optimized $\delta$ value for each galaxy ($E(B-V)_{\rm optimized}$) compared to those obtained with a Calzetti-like attenuation curve, i.e., $\delta=-1.0$, $E(B-V)_{\delta=-1.0}$, for galaxies not in the prohibited region in Figure \ref{fig:4fhafhb}. The top panel shows the absolute values, and the bottom panel shows their ratio, $\mathcal{R}$ = $E(B-V)_{\rm optimized}/E(B-V)_{\delta=-1.0}$. The two estimates can differ by up to $\sim$50\%, highlighting the importance of adopting an appropriate attenuation curve for accurate $E(B-V)$ estimation.}
    \label{fig:5ebvopt}
\end{figure}

With the optimized $\delta$ values for individual galaxies, we re-derived the $E(B-V)$ values and compared them with those obtained assuming an attenuation curve similar to the Calzetti law (i.e., $\delta=-1.0$, which gives $R_V=4.05$). 
Figure \ref{fig:5ebvopt} shows that the resulting $E(B-V)$ values can differ significantly, by $\sim10$--$40\%$. 
This highlights the importance of utilizing multiple hydrogen lines. 

Motivated by the diversity of attenuation-curve slopes inferred above, we further assessed how sensitive our Balmer-decrement-based $E(B-V)$ estimates are to the assumed attenuation curve. 
Specifically, for galaxies whose $E(B-V)$ values were derived in Section~\ref{sec:31dust} assuming the Calzetti law, we recalculated $E(B-V)$ using the SMC attenuation curve as well as a set of power-law attenuation curves with the $\delta$ values optimized for the eight galaxies discussed above. 
This test is not intended to optimize $\delta$ for individual galaxies, but rather to quantify the systematic uncertainty introduced by adopting a fixed attenuation law when only ${\rm H}\alpha$ and ${\rm H}\beta$ are available. 
We found that the estimated $E(B-V)$ values exhibit a modest but non-negligible systematic dependence on the adopted attenuation law. 
Specifically, adopting a steeper attenuation curve (i.e., a more negative $\delta$) systematically yields larger $E(B-V)$ values for a given Balmer decrement and broadens the overall $E(B-V)$ distribution, with the width increasing from 0.30 to 0.56.
The peak of the distribution shifts by at most 0.14 mag across the different curves.

While the wide range of observed Balmer decrements ($F_{\rm H\alpha}/F_{\rm H\beta}\sim$0.5--7) in our sample is the primary driver of the broad $E(B-V)$ distribution, this sensitivity to the assumed curve further emphasizes the importance of multi-line measurements for robust dust attenuation estimates.
In this context, the SPHEREx survey will provide a valuable database of such observations.
To explore this potential, we assess the detectability of additional hydrogen lines in our sample galaxies with SPHEREx by modeling mock SEDs, as described in the following section.

\section{Detectability of Hydrogen Lines with SPHEREx}\label{sec:4}
\subsection{SPHEREx mock SEDs with Paschen lines}\label{sec:41spherex}
\begin{figure}
    \centering
    \includegraphics[width=\linewidth]{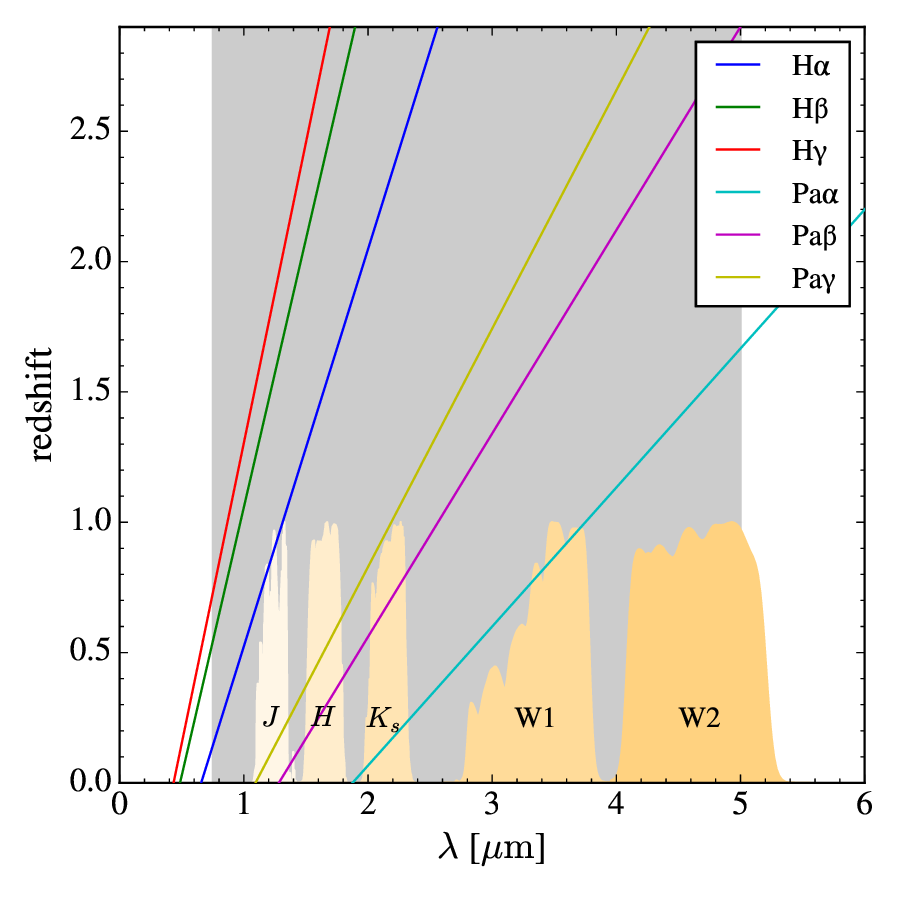}
    \caption{The observed wavelengths of hydrogen lines (solid lines in different colors) as a function of redshift, shown together with the SPHEREx wavelength coverage (gray shaded area) and the transmission curves of the 2MASS ($J$, $H$, and $K_s$) and catWISE (W1 and W2) photometric bands (orange shaded regions; transmission curves are from the SVO Filter Profile Service, \citealt{filter2012, filter2020, Rodrigo2024}). While the Balmer lines enter the SPHEREx wavelength coverage at $z\gtrsim0.2$, all three Paschen lines remain within the coverage up to $z\approx1.6$.}
    \label{fig:6spherex}
\end{figure}

In this section, we investigate the detectability of additional hydrogen lines in our sample galaxies with SPHEREx.
While only the 85 galaxies with detections of at least  two Balmer lines were analyzed in the previous section, we extend the sample to include 192 galaxies with spectroscopic redshifts, ${\rm Ly}\alpha$ fluxes, and catWISE photometry (W1 and/or W2) to examine the detectability of their Paschen lines with SPHEREx.
SPHEREx, with its infrared observation capabilities, provides an opportunity to detect the Balmer and/or Paschen lines of galaxies over a wide redshift range (Figure~\ref{fig:6spherex}); for our sample, the three Paschen lines (${\rm Pa}\alpha$, ${\rm Pa}\beta$ and ${\rm Pa}\gamma$) are fully covered by SPHEREx.

To construct SPHEREx mock SEDs with Paschen lines, we first modeled the continuum using an SED template from the Brown SED template library \citep{Browntemplate2014}.
Among the 129 galaxy SEDs, we selected one that closely matches the photometric SEDs of the three galaxies in our sample for which catWISE and 2MASS photometry are available.
The adopted template has a relatively steep optical-to-NIR continuum slope, which we consider more representative of star-forming galaxies than flatter alternatives in the Brown library.
We note that three galaxies with available catWISE and 2MASS photometry are too few to capture the full continuum diversity of the sample. 
Nevertheless, given the lack of comparable photometric constraints for the remaining galaxies, they provide the best available empirical basis for selecting a fiducial continuum template.
Consequently, the template is scaled to match the W1 magnitude of each galaxy, allowing us to construct individual continuum SEDs.

Each Paschen line, which is added on top of the continuum SED, is modeled as a Gaussian function. 
The line's total flux is derived from the extinction-corrected ${\rm H}\alpha$ or ${\rm Ly}\alpha$ line flux under the assumption of Case B recombination, i.e., $({\rm Pa}\alpha/{\rm H}\alpha)_{\rm int}=0.118$,
$({\rm Pa}\beta/{\rm H}\alpha)_{\rm int}=0.057$, and $({\rm Pa}\gamma/{\rm H}\alpha)_{\rm int}=0.031$. 
The extinction correction is done using either the estimated $E(B-V)$ for the ${\rm H}\alpha$ line or the typical value of $f_{\rm esc}({\rm Ly}\alpha)$ for our sample ($\sim10\%$) for the ${\rm Ly}\alpha$ line (see Section~\ref{sec:31dust}). 
When both ${\rm H}\alpha$ and ${\rm Ly}\alpha$ are available, ${\rm H}\alpha$ line is used to estimate the intrinsic strength of Paschen lines.
We then apply dust attenuation to the Paschen lines to predict their observed fluxes using the Calzetti attenuation curve. 
For galaxies lacking direct $E(B-V)$ measurements, no dust attenuation is applied to the predicted Paschen-line fluxes.
The line width is set to $400\,{\rm km\,s^{-1}}$, which is a typical value for star-forming galaxies \citep{Wisnioski2018, Calabr2022}. 
The line modeling is performed after converting the fluxes into units of ${\rm erg\,s^{-1}\,cm^{-2}\,\text{\AA}^{-1}}$.
The modeled SED (continuum plus lines) is then shifted to the observed redshift of each galaxy.

\begin{figure*}
\centering
    \includegraphics[width=\linewidth]{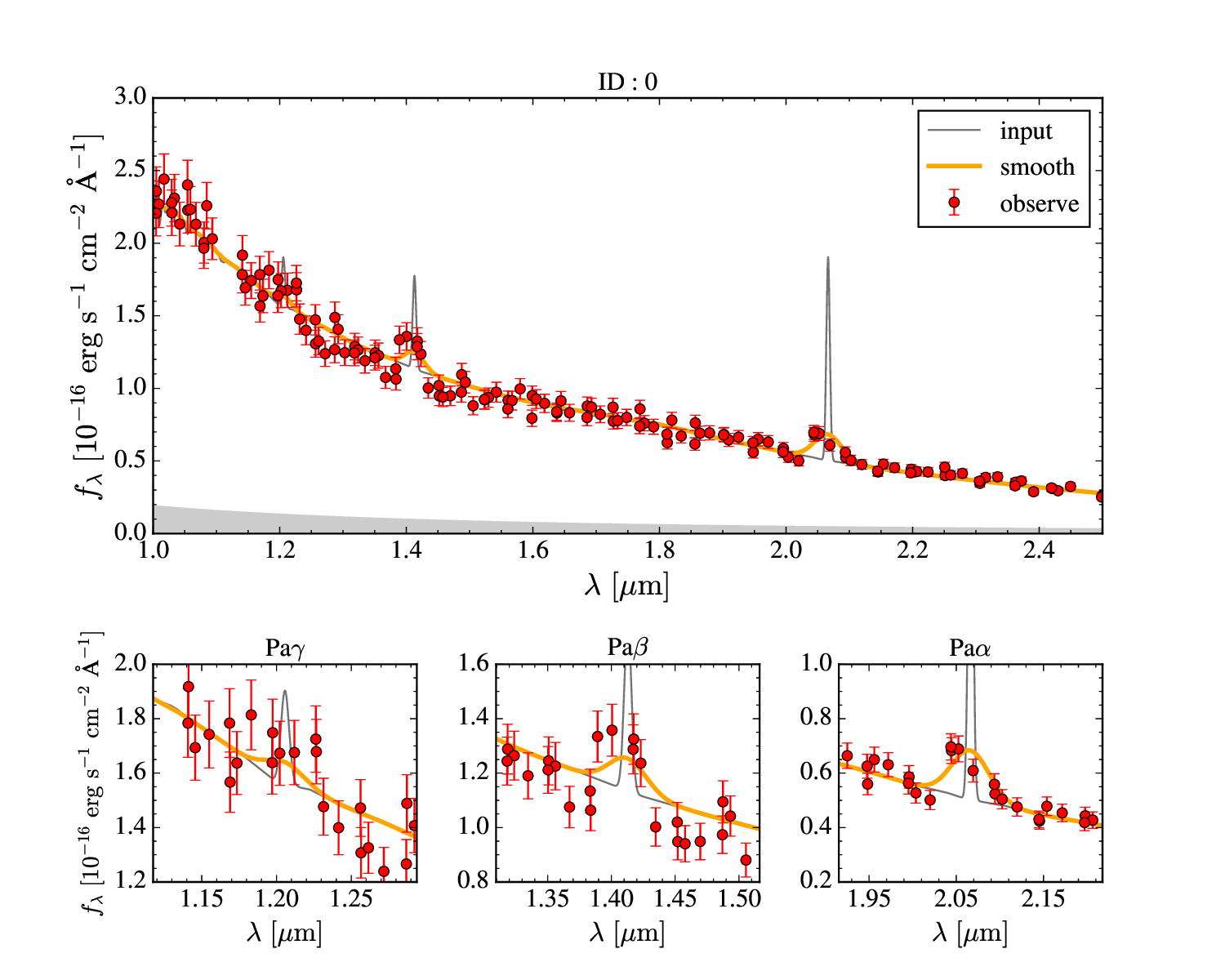}
    \caption{An example mock SPHEREx spectrum around the observed wavelength range of the Paschen lines for this particular galaxy (${\rm W1 = 14.79}$ mag) at $z=0.102$. The gray curve shows the modeled input spectrum to the SPHEREx simple simulator, the orange curve shows the spectrum smoothed to the SPHEREx resolution, and the red dots with error bars represent the simulated observed data including the SPHEREx photometric errors. The number of observed points is set to four times the number of SPHEREx channels to mimic the four repeated observations planned for the all-sky survey, without averaging nearby points to increase the signal-to-noise ratio, in order to better capture the line profiles with higher spectral resolution. The SPHEREx $5 \sigma$ detection limit is indicated by the gray shaded region. The bottom panels zoom in on the three Paschen lines, showing that while ${\rm Pa}\alpha$ and ${\rm Pa}\beta$ can be clearly identified, the ${\rm Pa}\gamma$ line is more challenging to detect.}
    \label{fig:7simulator}
\end{figure*}

To generate realistic mock SEDs, we used the SPHEREx simple simulator\footnote{\url{https://github.com/yongjungkim/spherexss}}.
The simulator operates through three main steps--smoothing, filtering, and observing. 
In the smoothing step, an input SED is smoothed according to the SPHEREx spectral resolution. 
In the filtering step, the smoothed SED is convolved with filter transmission curves, which are assumed to be a top-hat function with $\rm{100\%}$ efficiency for simplicity. 
In the observing step, the fluxes are converted into observed values by including photometric errors \citep{Ivezic2022}, implemented as Gaussian noise to perturb the fluxes. 
While the default setting provides 96 spectral channels\footnote{Although SPHEREx provides 102 spectral channels, the simple simulator uses the channel information provided in the SPHEREx Public Repository (\url{https://github.com/SPHEREx/Public-products}), which contains 96 channels.}, the four repeated observations planned for the all-sky survey over the two-year period will sample four times as many spectral channels. 
Since it is important to retain a sufficient number of data points to better capture spectral lines, rather than having fewer points with higher signal-to-noise, we configure the simulator to produce 96$\times$4 data points instead of 96.\footnote{The SPHEREx does not sample spectra at perfectly fixed wavelengths. 
To reflect this instrumental characteristic, we introduced a small random offset to each wavelength. Specifically, for each spectral point within a given channel, four wavelength points were generated by drawing from a uniform distribution covering the channel's wavelength range.}
Accordingly, the survey depth is adjusted to be shallower by $2.5\log_{10}(\sqrt{4}) \approx 0.75$ mag than the nominal survey depth (i.e., 19.4 for the all-sky survey), which is reached by averaging observed fluxes over nearby wavelengths obtained from repeated observations.
The mock SED generated using the simple simulator has been verified to be consistent with that produced by the simulator employed in the SPHEREx collaboration \citep[][accepted]{SPHERExsimulator2025}.

Figure \ref{fig:7simulator} shows an example of an input SED (gray curve), which is smoothed (orange) and then sampled to generate the mock observed data points (red dots with error bars). 
The zoomd-in views of each Paschen line are presented in the small panels at the bottom.
${\rm Pa}\alpha$ and ${\rm Pa}\beta$ are clearly detectable, whereas ${\rm Pa}\gamma$ appears challenging to observe in this case. 
In the following section, we describe the fitting procedure for individual lines, compare the measured line fluxes with the true input values, and evaluate how many of our sample galaxies would yield reliable Paschen line detections. 
It is worth noting that the expected number of galaxies with line detections is subject to uncertainties.
The main sources of these uncertainties are that the continuum SEDs are modeled using a single template, the SPHEREx filter response curves are approximated by top-hat functions, no dust attenuation--neither internal nor Milky Way--is applied to the SEDs, and the attenuation curve is assumed to follow Calzetti for all galaxies when deriving $E(B-V)$ to dust-correct ${\rm H}\alpha$ and compute the intrinsic fluxes of the Paschen lines.

\begin{figure*}
    \centering
    \includegraphics[width=\linewidth]{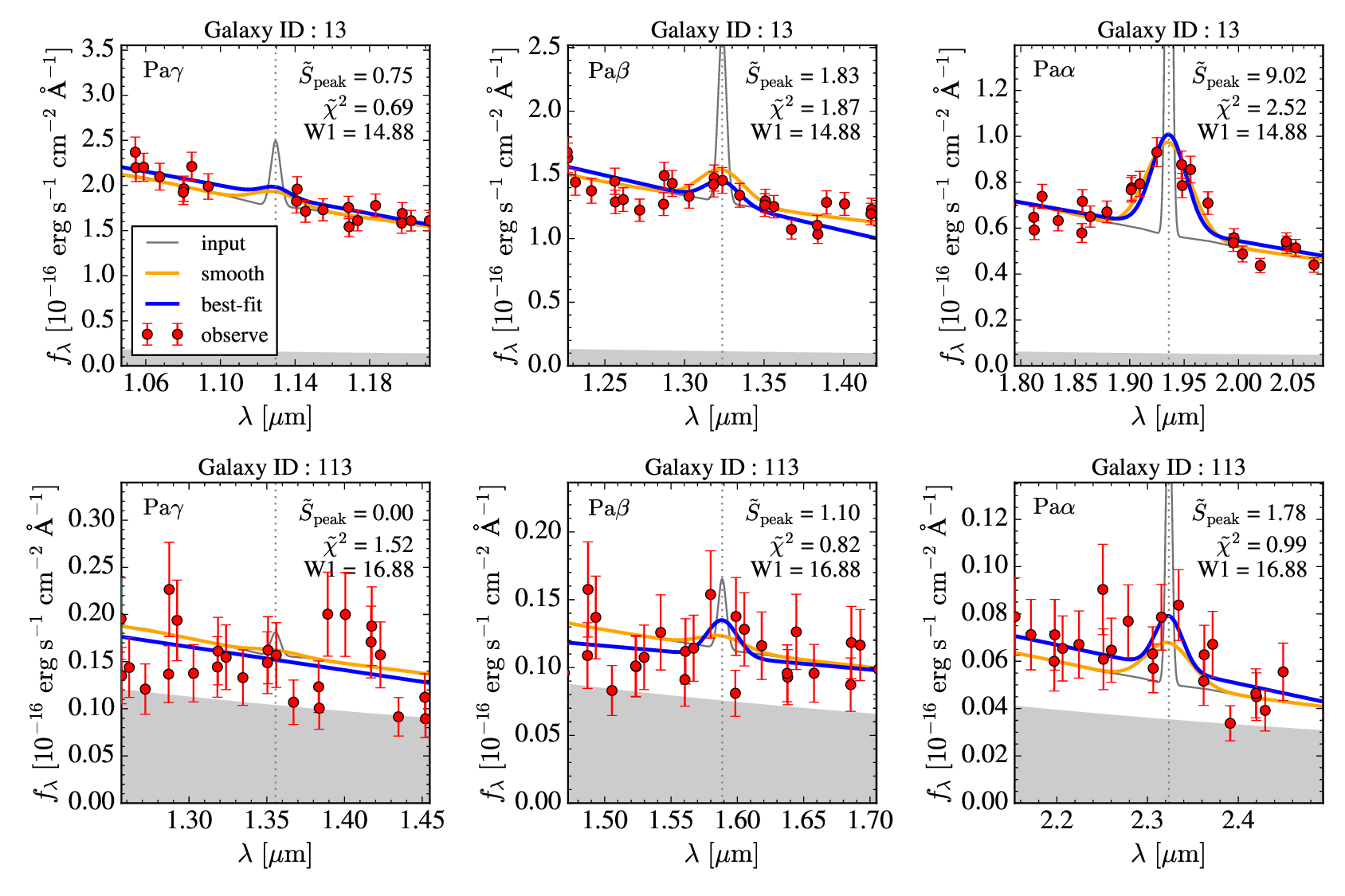}
    \caption{Examples of galaxies for which all three Paschen lines are detectable with SPHEREx (Galaxy IDs 13 and 113). The gray and orange curves, and the red dots with error bars are the same as in Figure \ref{fig:7simulator}, while the blue curves show the fitted spectra to the red points. The SPHEREx detection limit is indicated by the gray shades. The significance of each fitted line peak ($\tilde{S}_{\rm peak}$) is calculated relative to the mean error within each line window. The peak position, fixed during the fit based on the galaxy's redshift, is indicated by the dotted line. The reduced chi-square value ($\tilde{\chi}^2$) and W1 magnitude are provided in the legend along with $\tilde{S}_{\rm peak}$.}
    \label{fig:8fitexample}
\end{figure*}

\subsection{Measurement of Paschen lines in mock SEDs}\label{sec:42simulator}

\begin{figure*}
    \centering
    \includegraphics[width=\linewidth]{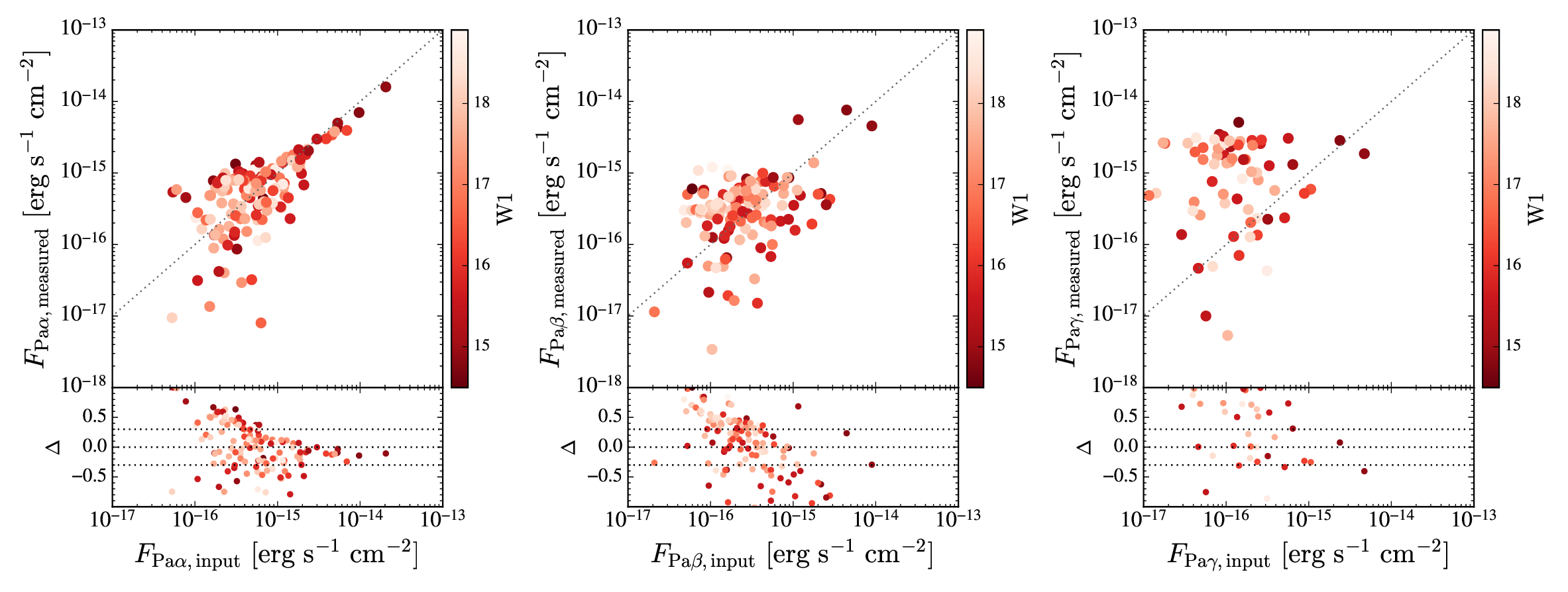}
    \caption{Measured line fluxes from the SPHEREx mock spectra compared to their corresponding true values for the ${\rm Pa}\alpha$ (left), ${\rm Pa}\beta$ (middle), and ${\rm Pa}\gamma$ (right) lines. There are fewer dots in the right panel for ${\rm Pa}\gamma$, as the measurement failed for many galaxies ($\sim67\%$) due to its weak emission. Dots are color-coded by the catWISE W1 magnitude (darker indicates brighter). The 1-1 relation is shown as a dotted line in each main panel. The bottom panels show the logarithmic differences ($\Delta = {\rm log} \, F_{{\rm measured}} - {\rm log} \, F_{{\rm input}}$) between the measured and true fluxes, where the scatter increases toward lower line fluxes. The $\pm$0.3 dex range is indicated by dotted horizontal lines along with the zero-difference line. Lines with fluxes $\gtrsim 2\times 10^{-15}\,{\rm erg\,s^{-1}\,cm^{-2}}$ appear to be detectable with $\lesssim$0.3 dex accuracy with SPHEREx.}
    \label{fig:9inputoutput}
\end{figure*}

In this section, we describe the fitting procedure applied to the SPHEREx mock SEDs of the Paschen lines generated in the previous section, employing Gaussian functions combined with a linear continuum component:
\begin{equation}
    A \exp{\left(-{\frac{(\lambda-\lambda_c)^2}{2\sigma_{\lambda}^2}}\right)} + B\lambda+C
\label{equ:5gaussian}
\end{equation}
where $A$, $\sigma_{\lambda}$, $B$, and $C$ are fitting parameters, with $\lambda_c$ being fixed to the central wavelength of each spectral line redshifted according to each galaxy's redshift.
$\lambda$, $\lambda_c$, and $\sigma_{\lambda}$ are given in units of $\mu$m.
The Gaussian fitting was performed using the \texttt{curve\_fit} function from Python \texttt{scipy} library.
In the fitting procedure, the parameters are explored within limited ranges or with initial guesses informed by observed points. 
The line window is set to a width of $\pm 3 \Delta \lambda$ from the central wavelength of the line, with $\Delta \lambda$ representing the spectral resolution.
Specifically, $\sigma_{\lambda}$ is constrained to be no smaller than 0.3 times the SPHEREx resolution, and the continuum slope and level are estimated from the points outside the line window.
This approach helps limit the parameter space and ensure that the fit closely follows the data.

Figure \ref{fig:8fitexample} shows the fitting results for two galaxies in which two or more Paschen lines are clearly identified and well fitted, as indicated by the reduced $\chi^2$ values. 
The significance of a line ($\tilde{S}_{\rm peak}$) is calculated as the height of the peak divided by the mean error within the line window.
While the ${\rm Pa}\alpha$ lines of these two galaxies are predicted to be detectable with high significance, their ${\rm Pa}\gamma$ lines are likely to be only marginally detectable.
The ${\rm Pa}\beta$ line of galaxy ID 113 is predicted to be clearly detectable, whereas it is unlikely to be detected in galaxy ID 13.
As ${\rm H}\alpha$ measurements are available for both galaxies, these additional Paschen line measurements will help constrain the detailed properties of dust and star formation.
More examples of line modeling and fitting results for selected galaxies with ${\rm W1} < 19$ mag are presented in Figures \ref{fig:a1fitalpha}, \ref{fig:a2fitbeta}, and \ref{fig:a3fitgamma} in the Appendix.

We conclude that Paschen lines are sufficiently identifiable given the detection limit and spectral resolution of SPHEREx for a significant fraction of our sample, particularly for ${\rm Pa}\alpha$, which is the strongest among the Paschen lines. 
In contrast, ${\rm Pa}\gamma$ is rarely detectable.
Furthermore, by examining how well the measured line fluxes reproduce the true fluxes, we assessed the reliability of the line detections. 
Figure~\ref{fig:9inputoutput} shows the measured versus true line fluxes of our sample galaxies, color-coded by their respective W1 magnitudes. 
The agreement between measured and true fluxes declines as line fluxes decrease; for line fluxes above $\sim2\times10^{-15}\,\mathrm{erg\,s^{-1}\,cm^{-2}}$, the measured line fluxes agree with the true values within $\sim$0.3 dex.
This corresponds to a SFR of $\sim20\,M_\odot{\rm\, yr^{-1}}$ at the median redshift of our sample, $z=0.26$.
Adopting a line flux threshold of $\sim2\times10^{-15}\,\mathrm{erg\,s^{-1}\,cm^{-2}}$ for reliable measurements results in fourteen galaxies for ${\rm Pa}\alpha$, of which six also have reliable ${\rm Pa}\beta$ measurements, and among these, two have reliable ${\rm Pa}\gamma$ measurements.
Two examples are shown in Figure~\ref{fig:8fitexample}.

Because the Paschen-line fluxes are not known a priori, it is useful to express the detectability in terms of previously measured quantities such as H$\alpha$ or Ly$\alpha$ line fluxes.
We therefore determine the flux thresholds above which 50\% of the galaxies are expected to yield reliable Pa$\alpha$ line flux measurements.
We consider the H$\alpha$- and Ly$\alpha$-based samples separately because the Paschen-line predictions are derived using different assumptions for the two subsets.
For H$\alpha$, this occurs at $F_{\mathrm{H}\alpha}\sim4\times10^{-15}\,\mathrm{erg\,s^{-1}\,cm^{-2}}$.
Among the 22 galaxies above this threshold, 11 are predicted to have reliable Pa$\alpha$ measurements.
Thus, selecting galaxies brighter than this H$\alpha$ flux yields a $\sim$50\% success rate for reliable Pa$\alpha$ detection.
Similarly, a Ly$\alpha$ flux threshold of $\sim9\times10^{-15}\,\mathrm{erg\,s^{-1}\,cm^{-2}}$ yields a $\sim$50\% success rate, with 8 of the 16 galaxies above this threshold predicted to have reliable Pa$\alpha$ measurements.
The Ly$\alpha$-based threshold should be regarded as substantially more uncertain than the H$\alpha$-based estimate, since it relies on uniform assumption of $f_{\rm esc}(\mathrm{Ly}\alpha)=10\%$ and does not incorporate individual dust attenuation estimates. It should therefore be interpreted as an approximate guideline rather than a robust predictive criterion.

To assess the robustness of these detectability estimates, we repeated the mock modeling under several alternative assumptions. 
Specifically, we considered two additional continuum SED templates from the Brown library, two additional emission-line widths around our fiducial value, a Gaussian filter response in place of the idealized top-hat filter, and alternative attenuation prescriptions including an SMC-like curve and a set of power-law attenuation curves with different slopes motivated by Section~\ref{sec:32constraints}. 
For each case, we compared the measured line fluxes with the true values and examined the resulting reliable-detection statistics.
We find that the distributions of the logarithmic differences between the measured and true fluxes and the numbers of reliably detected Paschen lines change only marginally across these tests.
In particular, the predicted numbers of detectable Pa$\alpha$, Pa$\beta$, and Pa$\gamma$ emitters remain broadly unchanged under the alternative assumptions considered here.

\section{Summary}\label{sec:5summary}
In this study, we investigated the potential of using multiple hydrogen recombination lines to constrain the dust attenuation of ${\rm Ly}\alpha$ emitters at relatively low redshifts ($z\lesssim0.4$).
We compiled a sample of 260 galaxies from the literature that have at least one hydrogen recombination line measurement. 
For the 85 galaxies with at least two Balmer lines, we estimated their $E(B-V)$ assuming Case B recombination and the Calzetti attenuation curve, and in turn ${\rm Ly}\alpha$ escape fractions were also derived.
For the 14 galaxies with all three Balmer lines, we optimized the extinction curve so that it yields consistent $E(B-V)$ estimates across the three available Balmer line ratios, demonstrating significant variety in the slope of the dust attenuation curve.
This highlights the advantage of utilizing multiple hydrogen lines to constrain dust attenuation more precisely.

As the majority of galaxies in our sample lack sufficient hydrogen line measurements to constrain dust attenuation, we evaluated the feasibility of detecting additional hydrogen lines in these galaxies with SPHEREx.
Our modeling of the Paschen lines in the sample suggests that SPHEREx will be capable of detecting these lines for a non-negligible fraction of star-forming galaxies, offering new opportunities to probe dust attenuation and star formation.
We note, however, that this modeling has several important limitations. 
First, our galaxy sample is compiled from six literature datasets with different selection functions, instruments, and observational depths, so it should not be regarded as a homogeneous statistical sample of LAEs. 
Second, our SPHEREx forecasting necessarily adopts simplified assumptions, including a fiducial continuum SED template, a fixed emission-line width, an idealized spectral response, and the Calzetti attenuation curve. 
Although the sensitivity tests presented in Section~\ref{sec:42simulator} show that the detectability statistics are only weakly affected by reasonable variations in these assumptions, the reported number of detectable objects should be interpreted as indicative forecasts rather than precise predictions for the full LAE population.

As SPHEREx observations are currently underway and data are gradually becoming available, a next step will be to compare our predictions with the actual observations to assess their accuracy.
If Paschen lines are indeed detected, we will combine these new measurements with existing hydrogen line data to more accurately constrain the $E(B-V)$, the slope of the dust attenuation curve, and the SFRs of these galaxies.


\acknowledgments
We would like to express our sincere gratitude to Minjin Kim, Dohyeong Kim, Yongjung Kim and Jeonghyun Pyo for their insightful comments and helpful discussions, which greatly improved the quality of this manuscript.
We also thank the anonymous referee for their constructive comments and suggestions, which improved the clarity of the paper.
This work was supported by the National Research Foundation of Korea (NRF), through grants funded by the Korean government (MSIT) (No. 2022R1A4A3031306).
This research was supported by Global - Learning \& Academic research Institution for Master's · PhD students, and Postdocs (G-LAMP) Program of the National Research Foundation of Korea (NRF) grant funded by the Ministry of Education (No. RS-2025-25442707). 
HS was supported by the NRF grant funded by the MSIT (No. RS-2024-00349364).
This research has made use of the SVO Filter Profile Service ``Carlos Rodrigo", funded by MCIN/AEI/10.13039/501100011033/ through grant PID2023-146210NB-I00.

\appendix\label{sec:appendix}

\section{Summary of the compiled galaxy properties}
We provide a table summarizing the compiled properties of the 260 low-$z$ LAEs in our sample, which include their basic information, available hydrogen line flux measurements, and our estimates of $E(B-V)$ and $f_{\rm esc}({\rm Ly}\alpha)$.
Table~\ref{tab:jkastable3} presents some example rows, and the full table will be made available in a machine-readable format in the accompanying material.

\renewcommand{\thetable}{A\arabic{table}}
\setcounter{table}{0}
\landscape
\begin{table}
\centering
\caption{Compiled properties of the 260 low-$z$ Ly$\alpha$ emitters\label{tab:jkastable3}}
\begin{threeparttable}
\begin{tabular}{rrrrrrrrrrrrrrr}
\toprule
ID & RA & DEC & Redshift & W1 & $F_{{\rm Ly}\alpha}$ & $F_{{\rm H}\alpha}$  & $F_{{\rm H}\beta}$   & $F_{{\rm H}\gamma}$ & $E(B-V)_{{\rm Calzetti}}$\tnote{$\ast$} & $\delta$\tnote{$\dagger$} & $E(B-V)_{\rm optimized}$\tnote{$\ddagger$} & $f_{\rm esc}({\rm Ly}\alpha)$ & Reference\tnote{\S} \\ 
\cmidrule(lr){6-9}
       &   $^{\circ}$   &   $^{\circ}$   &      &   ${\rm mag}$  & \multicolumn{4}{c}{$10^{-15}\,\mathrm{erg\,s^{-1}\,cm^{-2}}$} & &  &   &   $\%$   &              \\
\midrule
0 &  144.5562 &  54.4736 &  0.1021 & 14.79 &  67.41 &  64.50$\pm$ 0.66 &  20.00$\pm$ 0.25 &  8.46$\pm$ 0.15 & 0.10$\pm$0.01 & -4.27 &  0.17$\pm$ 0.02 & 8.77 & O14 \\ \addlinespace
1 &  195.4230 &  29.3814 &  0.0574 & 14.82 &  2.62 &  22.60$\pm$ 0.27 &  5.70$\pm$ 0.12 &  2.10$\pm$ 0.08 & 0.28$\pm$0.02 & -3.44 &  0.42$\pm$ 0.03 & 0.56 & O14 \\ \addlinespace
2 &  202.1836 &  43.9307 &  0.0280 & - &  484.89 &  207.00$\pm$ 2.39 &  66.80$\pm$ 0.79 &  30.90$\pm$ 0.46 & 0.07$\pm$0.01 & - & - & 21.82 & O14 \\ \addlinespace
3 &  209.9622 &  57.4397 &  0.0338 & - &  256.78 &  117.00$\pm$ 1.45 &  38.30$\pm$ 0.49 &  16.80$\pm$ 0.29 & 0.06$\pm$0.02 & -5.21 &  0.10$\pm$ 0.03 & 21.22 & O14 \\ \addlinespace
4 &  198.8867 &  62.1269 &  0.0311 & - &  90.86 &  265.00$\pm$ 3.07 &  44.40$\pm$ 0.52 &  15.00$\pm$ 0.27 & 0.63$\pm$0.01 & -0.82 &  0.66$\pm$ 0.01 & 0.57 & O14 \\ \addlinespace
14 &  215.8867 &  52.6237 &  0.2836 & 17.62 &  1.70 &  1.49$\pm$ 0.10 &  0.22$\pm$ 0.01 & - & 0.74$\pm$0.07 & - & - & 1.37 & S09 \\ \addlinespace
15 &  215.3526 &  52.6555 &  0.2582 & 15.81 &  1.80 &  2.58$\pm$ 0.10 &  0.74$\pm$ 0.07 & - & 0.17$\pm$0.09 & - & - & 4.72 & S09 \\ \addlinespace
16 &  215.1805 &  52.7188 &  0.2467 & 17.93 &  2.87 &  1.18$\pm$ 0.08 &  0.32$\pm$ 0.05 & - & 0.21$\pm$0.15 & - & - & 14.71 & S09 \\ \addlinespace
17 &  215.8241 &  52.7135 &  0.2770 & 15.63 &  2.62 &  2.58$\pm$ 0.13 & - & - & 0.00$\pm$0.04 & - & - & - & S09 \\ \addlinespace
52 &  190.0414 &  62.5617 &  0.2104 & 15.45 &  7.46 &  12.60 &  3.16 & - & 0.28$\pm$0.00 & - & - & 2.85 & C11 \\ \addlinespace
53 &  53.0019 & -28.1826 &  0.2795 & 17.41 &  2.21 & - & - & - & - & - & - & - & C11 \\ \addlinespace
54 &  214.7289 &  53.1300 &  0.2047 & - &  4.17 &  1.26 &  0.32 & - & 0.28$\pm$0.00 & - & - & 15.91 & C11 \\ \addlinespace
55 &  219.7461 &  34.9606 &  0.3740 & 16.16 &  3.64 &  0.50 & - & - & - & - & - & - & C11 \\ \addlinespace
129 &  214.9023 &  53.1600 &  0.2680 & 18.14 &  1.77 &  0.44$\pm$ 0.01 &  0.15$\pm$ 0.01 & - & 0.01$\pm$0.04 & - & - & 44.47 & F11 \\ \addlinespace
130 &  214.2079 &  52.8389 &  0.2828 & 17.61 &  1.71 &  0.29$\pm$ 0.02 &  0.06$\pm$ 0.01 & - & 0.40$\pm$0.12 & - & - & 19.52 & F11 \\ \addlinespace
143 &  10.0078 & -44.4288 &  0.2750 & 17.64 &  2.81 &  0.90$\pm$ 1.20 &  0.33$\pm$ 0.04 & - & -0.04$\pm$1.15 & - & - & - & A14 \\ \addlinespace
144 &  9.5590 & -44.2436 &  0.2780 & - &  17.90 &  2.39$\pm$ 0.10 &  0.73$\pm$ 0.03 & - & 0.12$\pm$0.05 & - & - & 60.38 & A14 \\ \addlinespace
145 &  9.8839 & -44.1917 &  0.1860 & - &  10.10 &  1.58$\pm$ 0.80 &  0.47$\pm$ 0.10 & - & 0.14$\pm$0.47 & - & - & 48.08 & A14 \\ \addlinespace
146 &  52.7601 & -27.8585 &  0.1853 & 15.77 &  8.63 & - & - & - & - & - & - & - & W17 \\ \addlinespace
147 &  52.7855 & -27.7041 &  0.2185 & - &  2.55 & - & - & - & - & - & - & - & W17 \\ \addlinespace
148 &  52.7971 & -27.8829 &  0.2509 & 18.09 &  1.70 & - & - & - & - & - & - & - & W17 \\ \addlinespace
\bottomrule
\end{tabular}
\begin{tablenotes}
\item[$\ast$] Calculated assuming the Calzetti attenuation curve. Negative values are unphysical, but are included as derived to reflect the measurement uncertainties and possible deviations from the Case B assumption. 
\item[$\dagger$] Optimized value by using all three Balmer lines when available (see Equations (\ref{equ:3alambda}) and (\ref{equ:4fhbfha})).
\item[$\ddagger$] Calculated using the optimized $\delta$ value. Galaxies with $\delta > 0$ are excluded from this calculation.
\item[\S] O14, S09, C11, F11, A14, and W17 refer to \citet{Ostlin2014}, \citet{Scarlata2009}, \citet{Cowie2011}, \citet{Finkelstein2011}, \citet{Atek2014}, and \citet{Wold2017}, respectively.
\end{tablenotes}
\end{threeparttable}
\label{tab:summary}
\end{table}
\endlandscape

\section{Examples of SPHEREx mock Paschen lines}
We present additional examples of SPHEREx mock Paschen lines, selected from four W1 magnitude bins, in Figures \ref{fig:a1fitalpha}, \ref{fig:a2fitbeta}, and \ref{fig:a3fitgamma}.

\renewcommand{\thefigure}{A\arabic{figure}}
\setcounter{figure}{0}
\begin{figure*}
    \centering
    \includegraphics[width=\linewidth]{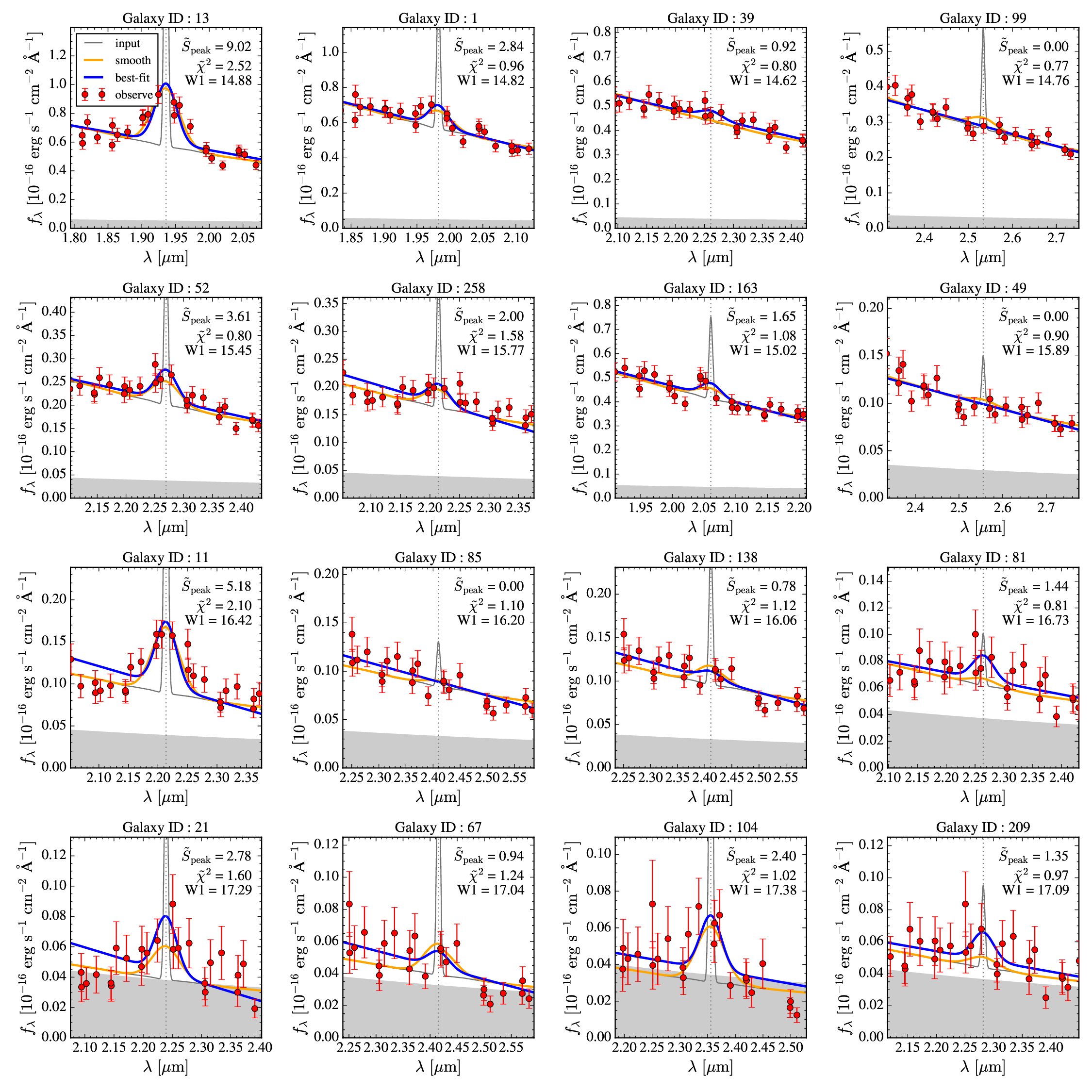}
    \caption{Example ${\rm Pa}\alpha$ lines, shown in the same format as Figure \ref{fig:8fitexample}. The gray, orange, and blue curves, and red dots represent the same quantities as in that figure. Each row shows galaxies grouped by their W1 magnitude. Toward the bottom rows (fainter magnitudes), the spectra approach the detection limit, indicated by the gray shades, and the error bars get larger. Various line strengths can be seen across the rows.}
    \label{fig:a1fitalpha}
\end{figure*}

\begin{figure*}
    \centering
    \includegraphics[width=0.5\textwidth]{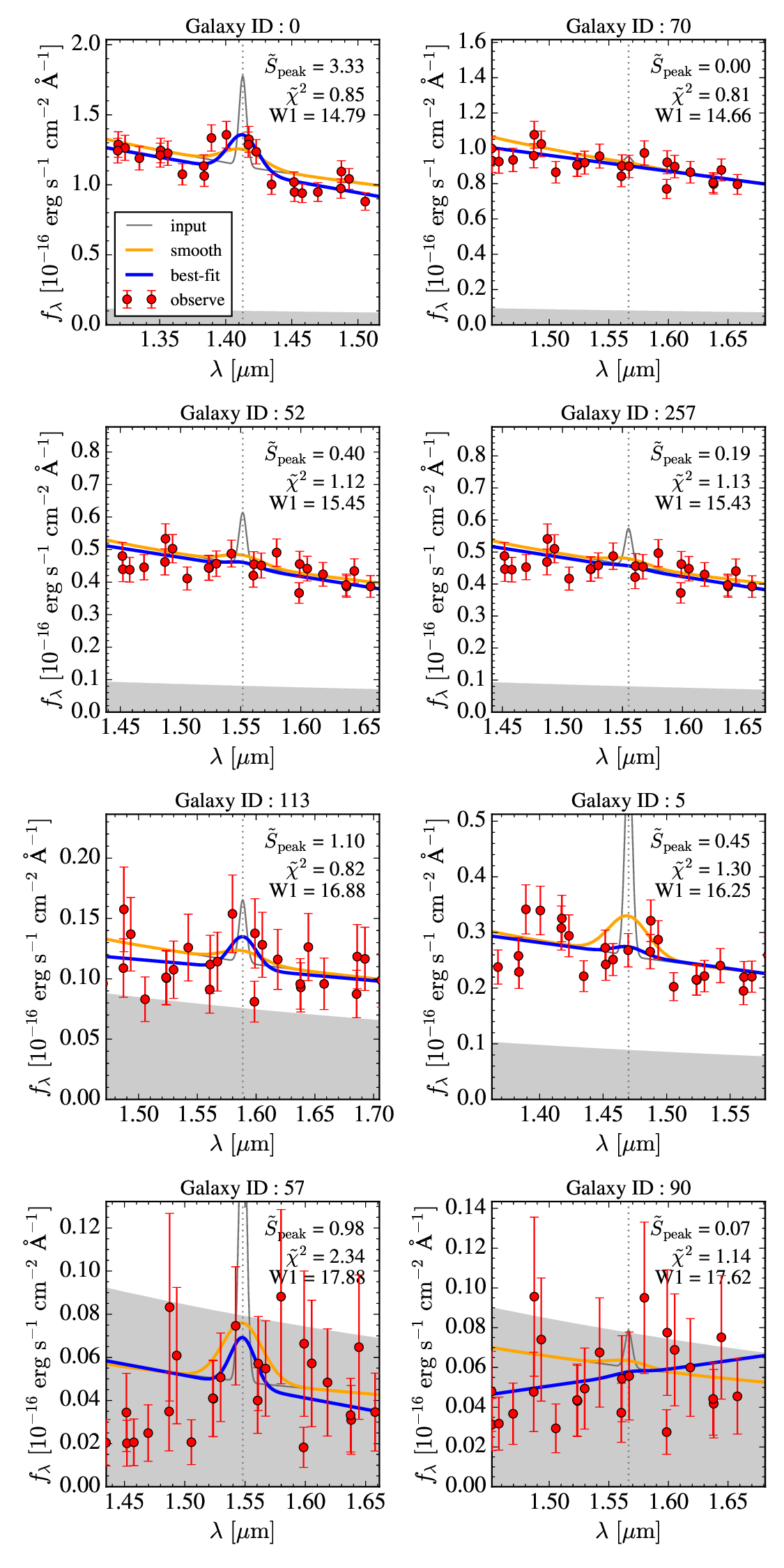}
    \caption{Similar to Figure \ref{fig:a1fitalpha}, but for the ${\rm Pa}\beta$ line.}
    \label{fig:a2fitbeta}
\end{figure*}

\begin{figure*}
    \centering
    \includegraphics[width=0.5\linewidth]{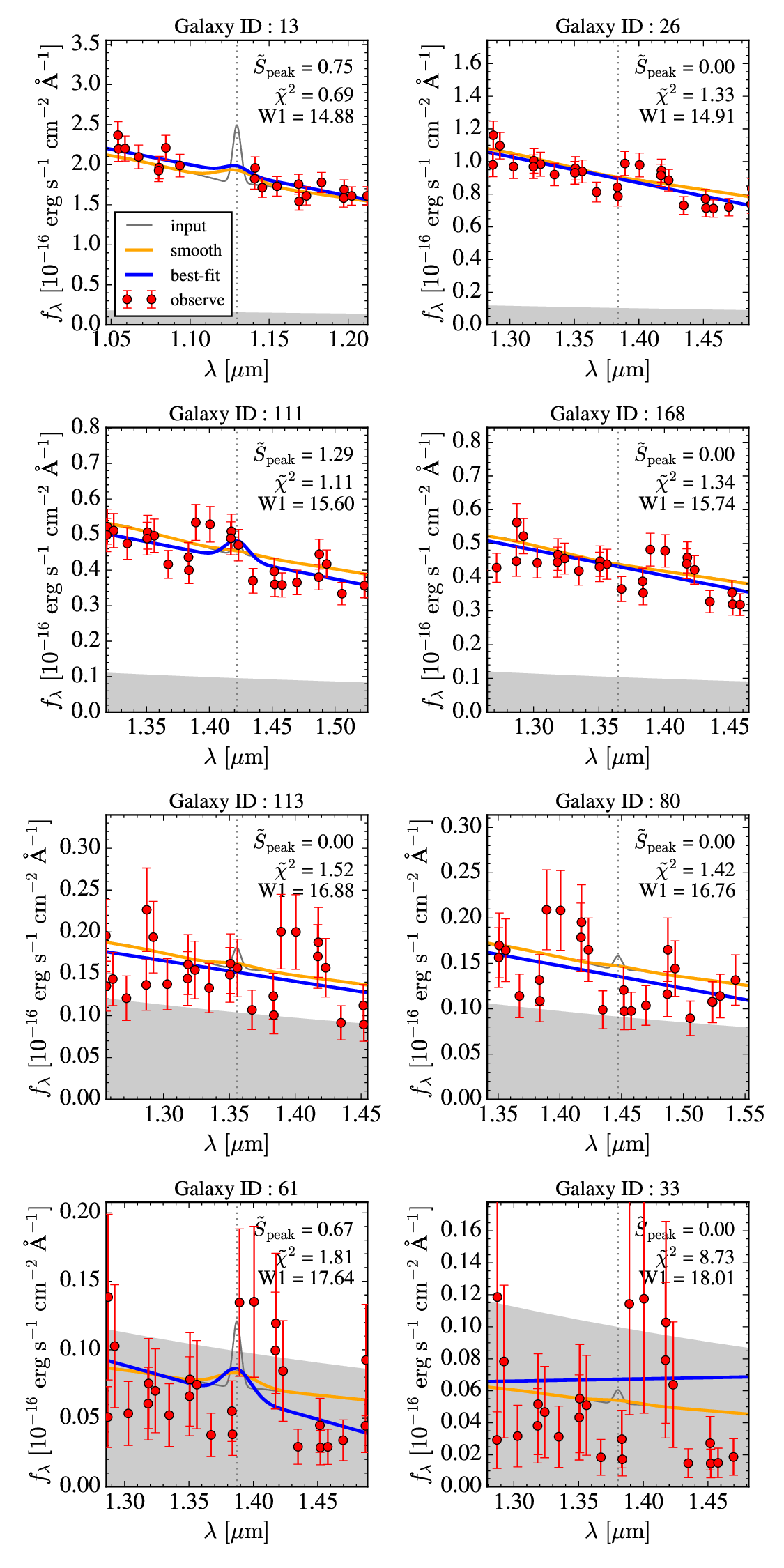}
    \caption{Similar to Figure \ref{fig:a1fitalpha}, but for the ${\rm Pa}\gamma$ line.}
    \label{fig:a3fitgamma}
\end{figure*}

\bibliography{ms}

\begin{thebibliography}{}
\expandafter\ifx\csname natexlab\endcsname\relax\def\natexlab#1{#1}\fi
\providecommand{\url}[1]{\href{#1}{#1}}
\providecommand{\dodoi}[1]{doi:~\href{http://doi.org/#1}{\nolinkurl{#1}}}
\providecommand{\doeprint}[1]{\href{http://ascl.net/#1}{\nolinkurl{http://ascl.net/#1}}}
\providecommand{\doarXiv}[1]{\href{https://arxiv.org/abs/#1}{\nolinkurl{https://arxiv.org/abs/#1}}}
\providecommand{\dodoilink}[2]{\href{http://doi.org/#1}{#2}}
\providecommand{\doadslink}[2]{\href{#1}{#2}}

\bibitem[{{Atek} {et~al.}(2014){Atek}, {Kunth}, {Schaerer}, {Mas-Hesse},
  {Hayes}, {{\"O}stlin}, \& {Kneib}}]{Atek2014}
{Atek}, H., {Kunth}, D., {Schaerer}, D., {et~al.} 2014, \aap, 561, A89

\bibitem[{{Brown} {et~al.}(2014){Brown}, {Moustakas}, {Smith}, {da Cunha},
  {Jarrett}, {Imanishi}, {Armus}, {Brandl}, \& {Peek}}]{Browntemplate2014}
{Brown}, M. J.~I., {Moustakas}, J., {Smith}, J. D.~T., {et~al.} 2014, \apjs,
  212, 18

\bibitem[{{Brown} {et~al.}(2017){Brown}, {Moustakas}, {Kennicutt}, {Bonne},
  {Intema}, {de Gasperin}, {Boquien}, {Jarrett}, {Cluver}, {Smith}, {da Cunha},
  {Imanishi}, {Armus}, {Brandl}, \& {Peek}}]{Brown2017}
{Brown}, M. J.~I., {Moustakas}, J., {Kennicutt}, R.~C., {et~al.} 2017, \apj,
  847, 136

\bibitem[{{Byun} {et~al.}(2021){Byun}, {Sheen}, {Seon}, {Ho}, {Lee}, {Jeong},
  {Kim}, {Park}, {Lee}, {Cha}, {Ko}, \& {Kim}}]{Byun2021}
{Byun}, W., {Sheen}, Y.-K., {Seon}, K.-I., {et~al.} 2021, \apj, 918, 82

\bibitem[{{Calabr{\`o}} {et~al.}(2022){Calabr{\`o}}, {Pentericci}, {Talia},
  {Cresci}, {Castellano}, {Belfiori}, {Mascia}, {Zamorani}, {Amor{\'\i}n},
  {Fynbo}, {Ginolfi}, {Guaita}, {Hathi}, {Koekemoer}, {Llerena}, {Mannucci},
  {Santini}, {Saxena}, \& {Schaerer}}]{Calabr2022}
{Calabr{\`o}}, A., {Pentericci}, L., {Talia}, M., {et~al.} 2022, \aap, 667,
  A117

\bibitem[{{Calzetti} {et~al.}(2000){Calzetti}, {Armus}, {Bohlin}, {Kinney},
  {Koornneef}, \& {Storchi-Bergmann}}]{Calzetti2000}
{Calzetti}, D., {Armus}, L., {Bohlin}, R.~C., {et~al.} 2000, \apj, 533, 682

\bibitem[{{Calzetti} {et~al.}(1994){Calzetti}, {Kinney}, \&
  {Storchi-Bergmann}}]{CalzettiKinney1994}
{Calzetti}, D., {Kinney}, A.~L., \& {Storchi-Bergmann}, T. 1994, \apj, 429, 582

\bibitem[{{Casey} {et~al.}(2014){Casey}, {Scoville}, {Sanders}, {Lee},
  {Cooray}, {Finkelstein}, {Capak}, {Conley}, {De Zotti}, {Farrah}, {Fu}, {Le
  Floc'h}, {Ilbert}, {Ivison}, \& {Takeuchi}}]{Casey2014}
{Casey}, C.~M., {Scoville}, N.~Z., {Sanders}, D.~B., {et~al.} 2014, \apj, 796,
  95

\bibitem[{{Cheng} {et~al.}(2020){Cheng}, {Ibar}, {Smail}, {Molina}, {Sobral},
  {Escala}, {Best}, {Cochrane}, {Gillman}, {Swinbank}, {Ivison}, {Huang},
  {Hughes}, {Villard}, \& {Cirasuolo}}]{Cheng2020}
{Cheng}, C., {Ibar}, E., {Smail}, I., {et~al.} 2020, \mnras, 499, 5241

\bibitem[{{Cleri} \& {Papovich}(2022)}]{Cleri2022}
{Cleri}, N., \& {Papovich}, C. 2022, in American Astronomical Society Meeting
  Abstracts, Vol.~54, American Astronomical Society Meeting Abstracts, 241.31

\bibitem[{{Cleri} {et~al.}(2022){Cleri}, {Trump}, {Backhaus}, {Momcheva},
  {Papovich}, {Simons}, {Weiner}, {Estrada-Carpenter}, {Finkelstein},
  {Giavalisco}, {Ji}, {Jung}, {Matharu}, {Martinez}, \&
  {Sturm}}]{CleriTrump2022}
{Cleri}, N.~J., {Trump}, J.~R., {Backhaus}, B.~E., {et~al.} 2022, \apj, 929, 3

\bibitem[{{Cochrane} {et~al.}(2017){Cochrane}, {Best}, {Sobral}, {Smail},
  {Wake}, {Stott}, \& {Geach}}]{Cochrane2017}
{Cochrane}, R.~K., {Best}, P.~N., {Sobral}, D., {et~al.} 2017, \mnras, 469,
  2913

\bibitem[{{Cowie} {et~al.}(2010){Cowie}, {Barger}, \& {Hu}}]{Cowie2010}
{Cowie}, L.~L., {Barger}, A.~J., \& {Hu}, E.~M. 2010, \apj, 711, 928

\bibitem[{{Cowie} {et~al.}(2011){Cowie}, {Barger}, \& {Hu}}]{Cowie2011}
{Cowie}, L.~L., {Barger}, A.~J., \& {Hu}, E.~M. 2011, \apj, 738, 136

\bibitem[{{Crill} {et~al.}(2020){Crill}, {Werner}, {Akeson}, {Ashby}, {Bleem},
  {Bock}, {Bryan}, {Burnham}, {Byunh}, {Chang}, {Chiang}, {Cook}, {Cooray},
  {Davis}, {Dor{\'e}}, {Dowell}, {Dubois-Felsmann}, {Eifler}, {Faisst},
  {Habib}, {Heinrich}, {Heitmann}, {Heaton}, {Hirata}, {Hristov}, {Hui},
  {Jeong}, {Kang}, {Kecman}, {Kirkpatrick}, {Korngut}, {Krause}, {Lee},
  {Lisse}, {Masters}, {Mauskopf}, {Melnick}, {Miyasaka}, {Nayyeri}, {Nguyen},
  {{\"O}berg}, {Padin}, {Paladini}, {Pourrahmani}, {Pyo}, {Smith}, {Song},
  {Symons}, {Teplitz}, {Tolls}, {Unwin}, {Windhorst}, {Yang}, \&
  {Zemcov}}]{SPHEREX2020}
{Crill}, B.~P., {Werner}, M., {Akeson}, R., {et~al.} 2020, in Society of
  Photo-Optical Instrumentation Engineers (SPIE) Conference Series, Vol. 11443,
  Space Telescopes and Instrumentation 2020: Optical, Infrared, and Millimeter
  Wave, ed. M.~{Lystrup} \& M.~D. {Perrin}, 114430I, \dodoi{10.1117/12.2567224}

\bibitem[{{Crill} {et~al.}(2025){Crill}, {Bach}, {Bryan}, {Choppin de Janvry},
  {Cukierman}, {Dowell}, {Everett}, {Fazar}, {Goldina}, {Huai}, {Hui}, {Jeong},
  {Kang}, {Korngut}, {Lee}, {Masters}, {Nguyen}, {Pyo}, {Symons}, {Yang},
  {Zemcov}, {Akeson}, {Ashby}, {Bock}, {Chang}, {Cheng}, {Chang}, {Cooray},
  {Dor{\'e}}, {Faisst}, {Feder}, \& {Werner}}]{SPHERExsimulator2025}
{Crill}, B.~P., {Bach}, Y.~P., {Bryan}, S.~A., {et~al.} 2025, arXiv e-prints,
  arXiv:2505.24856

\bibitem[{{Deharveng} {et~al.}(2008){Deharveng}, {Small}, {Barlow},
  {P{\'e}roux}, {Milliard}, {Friedman}, {Martin}, {Morrissey}, {Schiminovich},
  {Forster}, {Seibert}, {Wyder}, {Bianchi}, {Donas}, {Heckman}, {Lee},
  {Madore}, {Neff}, {Rich}, {Szalay}, {Welsh}, \& {Yi}}]{Deharveng2008Galex}
{Deharveng}, J.-M., {Small}, T., {Barlow}, T.~A., {et~al.} 2008, \apj, 680,
  1072

\bibitem[{{Finkelstein} {et~al.}(2011){Finkelstein}, {Cohen}, {Moustakas},
  {Malhotra}, {Rhoads}, \& {Papovich}}]{Finkelstein2011}
{Finkelstein}, S.~L., {Cohen}, S.~H., {Moustakas}, J., {et~al.} 2011, \apj,
  733, 117

\bibitem[{{Guo} {et~al.}(2016){Guo}, {Rafelski}, {Faber}, {Koo}, {Krumholz},
  {Trump}, {Willner}, {Amor{\'\i}n}, {Barro}, {Bell}, {Gardner}, {Gawiser},
  {Hathi}, {Koekemoer}, {Pacifici}, {P{\'e}rez-Gonz{\'a}lez}, {Ravindranath},
  {Reddy}, {Teplitz}, \& {Yesuf}}]{Guo2016}
{Guo}, Y., {Rafelski}, M., {Faber}, S.~M., {et~al.} 2016, \apj, 833, 37

\bibitem[{{Hamed} {et~al.}(2023){Hamed}, {Ma{\l}ek}, {Buat}, {Junais},
  {Ciesla}, {Donevski}, {Riccio}, \& {Figueira}}]{Hamed2023}
{Hamed}, M., {Ma{\l}ek}, K., {Buat}, V., {et~al.} 2023, \aap, 674, A99

\bibitem[{{Hayashi} {et~al.}(2013){Hayashi}, {Sobral}, {Best}, {Smail}, \&
  {Kodama}}]{HayashiSobral2013}
{Hayashi}, M., {Sobral}, D., {Best}, P.~N., {Smail}, I., \& {Kodama}, T. 2013,
  \mnras, 430, 1042

\bibitem[{{Herenz} {et~al.}(2019){Herenz}, {Wisotzki}, {Saust}, {Kerutt},
  {Urrutia}, {Diener}, {Schmidt}, {Marino}, {de la Vieuville}, {Boogaard},
  {Schaye}, {Guiderdoni}, {Richard}, \& {Bacon}}]{HerenzEdmund2019}
{Herenz}, E.~C., {Wisotzki}, L., {Saust}, R., {et~al.} 2019, \aap, 621, A107

\bibitem[{{Imanishi} {et~al.}(2010){Imanishi}, {Nakagawa}, {Shirahata},
  {Ohyama}, \& {Onaka}}]{Imanishi2010}
{Imanishi}, M., {Nakagawa}, T., {Shirahata}, M., {Ohyama}, Y., \& {Onaka}, T.
  2010, \apj, 721, 1233

\bibitem[{{Ivezi{\'c}} {et~al.}(2022){Ivezi{\'c}}, {Ivezi{\'c}}, {Moeyens},
  {Lisse}, {Bus}, {Jones}, {Crill}, {Dor{\'e}}, \& {Emery}}]{Ivezic2022}
{Ivezi{\'c}}, {\v{Z}}., {Ivezi{\'c}}, V., {Moeyens}, J., {et~al.} 2022,
  \icarus, 371, 114696

\bibitem[{{Ji} {et~al.}(2023){Ji}, {Yan}, {Bundy}, {Boquien}, {Schaefer},
  {Belfiore}, {Bershady}, {Drory}, {Li}, {Westfall}, {Lin}, {Bizyaev}, {Law},
  {Riffel}, \& {Riffel}}]{JiYan2023}
{Ji}, X., {Yan}, R., {Bundy}, K., {et~al.} 2023, \aap, 670, A125

\bibitem[{{Kennicutt}(1998)}]{Kennicutt1998}
{Kennicutt}, R. 1998, in ESA Special Publication, Vol. 429, LIA Colloq. 34: The
  Next Generation Space Telescope: Science Drivers and Technological
  Challenges, ed. B.~{Kaldeich-Sch{\"u}rmann}, 81,
  \dodoi{10.48550/arXiv.astro-ph/9807188}

\bibitem[{{Kennicutt} \& {Evans}(2012)}]{KennicuttEvans2012}
{Kennicutt}, R.~C., \& {Evans}, N.~J. 2012, \araa, 50, 531

\bibitem[{{Khostovan} {et~al.}(2024){Khostovan}, {Malhotra}, {Rhoads},
  {Sobral}, {Harish}, {Tilvi}, {Coughlin}, \& {Rezaee}}]{Khostovan2024}
{Khostovan}, A.~A., {Malhotra}, S., {Rhoads}, J.~E., {et~al.} 2024, arXiv
  e-prints, arXiv:2408.00080

\bibitem[{{Kim} \& {Im}(2018)}]{KimIm2018}
{Kim}, D., \& {Im}, M. 2018, \aap, 610, A31

\bibitem[{{Kim} {et~al.}(2010){Kim}, {Im}, \& {Kim}}]{Kim2010}
{Kim}, D., {Im}, M., \& {Kim}, M. 2010, \apj, 724, 386

\bibitem[{{Kim} {et~al.}(2023){Kim}, {Im}, {Kim}, {Kim}, {Shin}, {Shim}, \&
  {Song}}]{Kim2023quasar}
{Kim}, D., {Im}, M., {Kim}, M., {et~al.} 2023, \apj, 954, 156

\bibitem[{{Kim} {et~al.}(2018){Kim}, {Im}, {Canalizo}, {Kim}, {Kim}, {Woo},
  {Taak}, {Kim}, \& {Lazarova}}]{Kim2018ebv}
{Kim}, D., {Im}, M., {Canalizo}, G., {et~al.} 2018, \apjs, 238, 37

\bibitem[{{Lee} {et~al.}(2024){Lee}, {Gawiser}, {Park}, {Yang}, {Valdes},
  {Lang}, {Ramakrishnan}, {Moon}, {Firestone}, {Appleby}, {Artale}, {Andrews},
  {Bauer}, {Benda}, {Broussard}, {Chiang}, {Ciardullo}, {Dey}, {Farooq},
  {Gronwall}, {Guaita}, {Huang}, {Hwang}, {Im}, {Jeong}, {Karthikeyan}, {Kim},
  {Kim}, {Kumar}, {Nagaraj}, {Nantais}, {Padilla}, {Park}, {Pope}, {Popescu},
  {Schlegel}, {Seo}, {Singh}, {Song}, {Troncoso}, {Vivas}, {Zabludoff}, \&
  {Zenteno}}]{ODIN}
{Lee}, K.-S., {Gawiser}, E., {Park}, C., {et~al.} 2024, \apj, 962, 36

\bibitem[{{Lin} \& {Yan}(2024)}]{LinYan2024}
{Lin}, Z., \& {Yan}, R. 2024, \aap, 691, A201

\bibitem[{{Mentuch Cooper} {et~al.}(2023){Mentuch Cooper}, {Gebhardt}, {Davis},
  {Farrow}, {Liu}, {Zeimann}, {Ciardullo}, {Feldmeier}, {Drory}, {Jeong},
  {Benda}, {Bowman}, {Boylan-Kolchin}, {Ch{\'a}vez Ortiz}, {Debski}, {Dentler},
  {Fabricius}, {Farooq}, {Finkelstein}, {Gawiser}, {Gronwall}, {Hill}, {Hopp},
  {House}, {Janowiecki}, {Khoraminezhad}, {Kollatschny}, {Komatsu}, {Landriau},
  {Niemeyer}, {Lee}, {MacQueen}, {Mawatari}, {McKay}, {Ouchi}, {Poppe},
  {Saito}, {Schneider}, {Snigula}, {Thomas}, {Tuttle}, {Urrutia}, {Weiss},
  {Wisotzki}, {Zhang}, \& {HETDEX Collaboration}}]{Cooper2023}
{Mentuch Cooper}, E., {Gebhardt}, K., {Davis}, D., {et~al.} 2023, \apj, 943,
  177

\bibitem[{{Meurer} {et~al.}(1999){Meurer}, {Heckman}, \&
  {Calzetti}}]{Meurer1999}
{Meurer}, G.~R., {Heckman}, T.~M., \& {Calzetti}, D. 1999, in American
  Institute of Physics Conference Series, Vol. 470, After the Dark Ages: When
  Galaxies were Young (the Universe at 2 < Z < 5), ed. S.~{Holt} \&
  E.~{Smith}, 359--363, \dodoi{10.1063/1.58622}

\bibitem[{{Osterbrock} \& {Ferland}(2006)}]{OsterbrockFerland2006}
{Osterbrock}, D.~E., \& {Ferland}, G.~J. 2006, {Astrophysics of gaseous nebulae
  and active galactic nuclei}

\bibitem[{{{\"O}stlin} {et~al.}(2014){{\"O}stlin}, {Hayes}, {Duval},
  {Sandberg}, {Rivera-Thorsen}, {Marquart}, {Orlitov{\'a}}, {Adamo},
  {Melinder}, {Guaita}, {Atek}, {Cannon}, {Gruyters}, {Herenz}, {Kunth},
  {Laursen}, {Mas-Hesse}, {Micheva}, {Ot{\'\i}-Floranes}, {Pardy}, {Roth},
  {Schaerer}, \& {Verhamme}}]{Ostlin2014}
{{\"O}stlin}, G., {Hayes}, M., {Duval}, F., {et~al.} 2014, \apj, 797, 11

\bibitem[{{Ouchi} {et~al.}(2008){Ouchi}, {Shimasaku}, {Akiyama}, {Simpson},
  {Saito}, {Ueda}, {Furusawa}, {Sekiguchi}, {Yamada}, {Kodama}, {Kashikawa},
  {Okamura}, {Iye}, {Takata}, {Yoshida}, \& {Yoshida}}]{Ouchi2008}
{Ouchi}, M., {Shimasaku}, K., {Akiyama}, M., {et~al.} 2008, \apjs, 176, 301

\bibitem[{{Ouchi} {et~al.}(2018){Ouchi}, {Harikane}, {Shibuya}, {Shimasaku},
  {Taniguchi}, {Konno}, {Kobayashi}, {Kajisawa}, {Nagao}, {Ono}, {Inoue},
  {Umemura}, {Mori}, {Hasegawa}, {Higuchi}, {Komiyama}, {Matsuda}, {Nakajima},
  {Saito}, \& {Wang}}]{Ouchi2018}
{Ouchi}, M., {Harikane}, Y., {Shibuya}, T., {et~al.} 2018, \pasj, 70, S13

\bibitem[{Pastrav(2023)}]{Pastrav2023}
Pastrav, B.~A. 2023, Monthly Notices of the Royal Astronomical Society, 527,
  11167

\bibitem[{{Payne} {et~al.}(2018){Payne}, {Inami}, {Malkan}, {Matsuhara}, \&
  {Sakai}}]{Payne2018}
{Payne}, A.~V., {Inami}, H., {Malkan}, M.~A., {Matsuhara}, H., \& {Sakai}, S.
  2018, in The Cosmic Wheel and the Legacy of the AKARI Archive: From Galaxies
  and Stars to Planets and Life, ed. T.~{Ootsubo}, I.~{Yamamura}, K.~{Murata},
  \& T.~{Onaka}, 337--340

\bibitem[{{Pettini} {et~al.}(2001){Pettini}, {Shapley}, {Steidel}, {Cuby},
  {Dickinson}, {Moorwood}, {Adelberger}, \& {Giavalisco}}]{PettiniShapley2001}
{Pettini}, M., {Shapley}, A.~E., {Steidel}, C.~C., {et~al.} 2001, \apj, 554,
  981

\bibitem[{{Planck Collaboration} {et~al.}(2020){Planck Collaboration},
  {Aghanim}, {Akrami}, {Ashdown}, {Aumont}, {Baccigalupi}, {Ballardini},
  {Banday}, {Barreiro}, {Bartolo}, {Basak}, {Battye}, {Benabed}, {Bernard},
  {Bersanelli}, {Bielewicz}, {Bock}, {Bond}, {Borrill}, {Bouchet}, {Boulanger},
  {Bucher}, {Burigana}, {Butler}, {Calabrese}, {Cardoso}, {Carron},
  {Challinor}, {Chiang}, {Chluba}, {Colombo}, {Combet}, {Contreras}, {Crill},
  {Cuttaia}, {de Bernardis}, {de Zotti}, {Delabrouille}, {Delouis}, {Di
  Valentino}, {Diego}, {Dor{\'e}}, {Douspis}, {Ducout}, {Dupac}, {Dusini},
  {Efstathiou}, {Elsner}, {En{\ss}lin}, {Eriksen}, {Fantaye}, {Farhang},
  {Fergusson}, {Fernandez-Cobos}, {Finelli}, {Forastieri}, {Frailis},
  {Fraisse}, {Franceschi}, {Frolov}, {Galeotta}, {Galli}, {Ganga},
  {G{\'e}nova-Santos}, {Gerbino}, {Ghosh}, {Gonz{\'a}lez-Nuevo}, {G{\'o}rski},
  {Gratton}, {Gruppuso}, {Gudmundsson}, {Hamann}, {Handley}, {Hansen},
  {Herranz}, {Hildebrandt}, {Hivon}, {Huang}, {Jaffe}, {Jones}, {Karakci},
  {Keih{\"a}nen}, {Keskitalo}, {Kiiveri}, {Kim}, {Kisner}, {Knox},
  {Krachmalnicoff}, {Kunz}, {Kurki-Suonio}, {Lagache}, {Lamarre}, {Lasenby},
  {Lattanzi}, {Lawrence}, {Le Jeune}, {Lemos}, {Lesgourgues}, {Levrier},
  {Lewis}, {Liguori}, {Lilje}, {Lilley}, {Lindholm}, {L{\'o}pez-Caniego},
  {Lubin}, {Ma}, {Mac{\'\i}as-P{\'e}rez}, {Maggio}, {Maino}, {Mandolesi},
  {Mangilli}, {Marcos-Caballero}, {Maris}, {Martin}, {Martinelli},
  {Mart{\'\i}nez-Gonz{\'a}lez}, {Matarrese}, {Mauri}, {McEwen}, {Meinhold},
  {Melchiorri}, {Mennella}, {Migliaccio}, {Millea}, {Mitra},
  {Miville-Desch{\^e}nes}, {Molinari}, {Montier}, {Morgante}, {Moss}, {Natoli},
  {N{\o}rgaard-Nielsen}, {Pagano}, {Paoletti}, {Partridge}, {Patanchon},
  {Peiris}, {Perrotta}, {Pettorino}, {Piacentini}, {Polastri}, {Polenta},
  {Puget}, {Rachen}, {Reinecke}, {Remazeilles}, {Renzi}, {Rocha}, {Rosset},
  {Roudier}, {Rubi{\~n}o-Mart{\'\i}n}, {Ruiz-Granados}, {Salvati}, {Sandri},
  {Savelainen}, {Scott}, {Shellard}, {Sirignano}, {Sirri}, {Spencer},
  {Sunyaev}, {Suur-Uski}, {Tauber}, {Tavagnacco}, {Tenti}, {Toffolatti},
  {Tomasi}, {Trombetti}, {Valenziano}, {Valiviita}, {Van Tent}, {Vibert},
  {Vielva}, {Villa}, {Vittorio}, {Wandelt}, {Wehus}, {White}, {White},
  {Zacchei}, \& {Zonca}}]{PlankCollaboration2020}
{Planck Collaboration}, {Aghanim}, N., {Akrami}, Y., {et~al.} 2020, \aap, 641,
  A6

\bibitem[{{Prescott} {et~al.}(2022){Prescott}, {Finlator}, {Cleri}, {Trump}, \&
  {Papovich}}]{Prescott2022}
{Prescott}, M. K.~M., {Finlator}, K.~M., {Cleri}, N.~J., {Trump}, J.~R., \&
  {Papovich}, C. 2022, \apj, 928, 71

\bibitem[{{Reddy} {et~al.}(2023){Reddy}, {Topping}, {Sanders}, {Shapley}, \&
  {Brammer}}]{Reddy2023}
{Reddy}, N.~A., {Topping}, M.~W., {Sanders}, R.~L., {Shapley}, A.~E., \&
  {Brammer}, G. 2023, \apj, 948, 83

\bibitem[{{Rodrigo} \& {Solano}(2020)}]{filter2020}
{Rodrigo}, C., \& {Solano}, E. 2020, in XIV.0 Scientific Meeting (virtual) of
  the Spanish Astronomical Society, 182

\bibitem[{{Rodrigo} {et~al.}(2012){Rodrigo}, {Solano}, \& {Bayo}}]{filter2012}
{Rodrigo}, C., {Solano}, E., \& {Bayo}, A. 2012, {SVO Filter Profile Service
  Version 1.0}, IVOA Working Draft 15 October 2012,
  \dodoi{10.5479/ADS/bib/2012ivoa.rept.1015R}

\bibitem[{{Rodrigo} {et~al.}(2024){Rodrigo}, {Cruz}, {Aguilar}, {Aller},
  {Solano}, {G{\'a}lvez-Ortiz}, {Jim{\'e}nez-Esteban}, {Mas-Buitrago}, {Bayo},
  {Cort{\'e}s-Contreras}, {Murillo-Ojeda}, {Bonoli}, {Cenarro}, {Dupke},
  {L{\'o}pez-Sanjuan}, {Mar{\'\i}n-Franch}, {de Oliveira}, {Moles}, {Taylor},
  {Varela}, \& {Rami{\'o}}}]{Rodrigo2024}
{Rodrigo}, C., {Cruz}, P., {Aguilar}, J.~F., {et~al.} 2024, \aap, 689, A93

\bibitem[{{Salim} \& {Narayanan}(2020)}]{Salimnaraynan2020}
{Salim}, S., \& {Narayanan}, D. 2020, \araa, 58, 529

\bibitem[{{Scarlata} {et~al.}(2024){Scarlata}, {Hayes}, {Panagia}, {Mehta},
  {Haardt}, \& {Bagley}}]{Scarlate2024}
{Scarlata}, C., {Hayes}, M., {Panagia}, N., {et~al.} 2024, arXiv e-prints,
  arXiv:2404.09015

\bibitem[{{Scarlata} {et~al.}(2009){Scarlata}, {Colbert}, {Teplitz}, {Panagia},
  {Hayes}, {Siana}, {Rau}, {Francis}, {Caon}, {Pizzella}, \&
  {Bridge}}]{Scarlata2009}
{Scarlata}, C., {Colbert}, J., {Teplitz}, H.~I., {et~al.} 2009, \apjl, 704, L98

\bibitem[{Sun {et~al.}(2018)Sun, Greene, Zakamska, Goulding, Strauss, Huang,
  Johnson, Kawaguchi, Matsuoka, Marsteller, Nagao, \& Toba}]{Sun2018}
Sun, A.-L., Greene, J.~E., Zakamska, N.~L., {et~al.} 2018, Monthly Notices of
  the Royal Astronomical Society, 480, 2302

\bibitem[{{Takeuchi} {et~al.}(2012){Takeuchi}, {Yuan}, {Ikeyama}, {Murata}, \&
  {Inoue}}]{Takeuhi2012}
{Takeuchi}, T.~T., {Yuan}, F.-T., {Ikeyama}, A., {Murata}, K.~L., \& {Inoue},
  A.~K. 2012, \apj, 755, 144

\bibitem[{{Whitaker} {et~al.}(2014){Whitaker}, {Franx}, {Leja}, {van Dokkum},
  {Henry}, {Skelton}, {Fumagalli}, {Momcheva}, {Brammer}, {Labb{\'e}},
  {Nelson}, \& {Rigby}}]{Whitaker2014}
{Whitaker}, K.~E., {Franx}, M., {Leja}, J., {et~al.} 2014, \apj, 795, 104

\bibitem[{{Wisnioski} {et~al.}(2018){Wisnioski}, {Mendel}, {F{\"o}rster
  Schreiber}, {Genzel}, {Wilman}, {Wuyts}, {Belli}, {Beifiori}, {Bender},
  {Brammer}, {Chan}, {Davies}, {Davies}, {Fabricius}, {Fossati}, {Galametz},
  {Lang}, {Lutz}, {Nelson}, {Momcheva}, {Rosario}, {Saglia}, {Tacconi},
  {Tadaki}, {{\"U}bler}, \& {van Dokkum}}]{Wisnioski2018}
{Wisnioski}, E., {Mendel}, J.~T., {F{\"o}rster Schreiber}, N.~M., {et~al.}
  2018, \apj, 855, 97

\bibitem[{{Wold} {et~al.}(2017){Wold}, {Finkelstein}, {Barger}, {Cowie}, \&
  {Rosenwasser}}]{Wold2017}
{Wold}, I. G.~B., {Finkelstein}, S.~L., {Barger}, A.~J., {Cowie}, L.~L., \&
  {Rosenwasser}, B. 2017, \apj, 848, 108

\bibitem[{{Wuyts} {et~al.}(2011){Wuyts}, {F{\"o}rster Schreiber}, {van der
  Wel}, {Magnelli}, {Guo}, {Genzel}, {Lutz}, {Aussel}, {Barro}, {Berta},
  {Cava}, {Graci{\'a}-Carpio}, {Hathi}, {Huang}, {Kocevski}, {Koekemoer},
  {Lee}, {Le Floc'h}, {McGrath}, {Nordon}, {Popesso}, {Pozzi}, {Riguccini},
  {Rodighiero}, {Saintonge}, \& {Tacconi}}]{Wuyts2011}
{Wuyts}, S., {F{\"o}rster Schreiber}, N.~M., {van der Wel}, A., {et~al.} 2011,
  \apj, 742, 96

\bibitem[{{Yano} {et~al.}(2021){Yano}, {Baba}, {Nakagawa}, {Malkan}, {Isobe},
  {Shirahata}, {Doi}, \& {Bhalotia}}]{Yano2021}
{Yano}, K., {Baba}, S., {Nakagawa}, T., {et~al.} 2021, \apj, 922, 272

\end{thebibliography}




\end{document}